\documentclass[twoside]{article}
\usepackage{qic}

\usepackage{amsmath}
\usepackage{graphicx}
\usepackage{xcolor}

\let\captionbox\relax
\usepackage{subcaption}

\begin{document}
\setlength{\textheight}{8.0truein}

\runninghead{A Cost-Effective Layout-Aware Quantum Circuit Synthesis
For Triangular, Square, and Heavy-Hex Layouts  $\ldots$}
{Sonia Yang, Ali Al-Bayaty, and Marek Perkowski $\ldots$}

\normalsize\textlineskip
\thispagestyle{empty}
\setcounter{page}{1}

% \copyrightheading{Vol.}{No.}{Year.}{000--000}

\vspace*{0.88truein}

\alphfootnote

\fpage{1}

\centerline{\bf A COST-EFFECTIVE LAYOUT-AWARE QUANTUM CIRCUIT SYNTHESIS}
\vspace*{0.035truein}
\centerline{\bf ON TRIANGULAR, SQUARE, AND HEAVY-HEX LAYOUTS}
\vspace*{0.37truein}

\centerline{\footnotesize
SONIA YANG\footnote{sonia.liu.yang@gmail.com}}
\vspace*{0.015truein}
\centerline{\footnotesize\it Department of Electrical and Computer Engineering, Portland State University}
\baselineskip=10pt
\centerline{\footnotesize\it Portland, Oregon 97201, United States}
\vspace*{10pt}

\centerline{\footnotesize
ALI AL-BAYATY\footnote{albayaty@pdx.edu}}
\vspace*{0.015truein}
\centerline{\footnotesize\it Department of Electrical and Computer Engineering, Portland State University}
\baselineskip=10pt
\centerline{\footnotesize\it Portland, Oregon 97201, United States}
\vspace*{10pt}

\centerline{\footnotesize
MAREK PERKOWSKI\footnote{h8mp@pdx.edu}}
\vspace*{0.015truein}
\centerline{\footnotesize\it Department of Electrical and Computer Engineering, Portland State University}
\baselineskip=10pt
\centerline{\footnotesize\it Portland, Oregon 97201, United States}
\vspace*{10pt}

\publisher{(received date)}{(revised date)}

\vspace*{0.21truein}

% Abstract
\abstracts{
The quantum layout and the mapping of logical to physical qubits are crucial in quantum circuit synthesis for a real quantum computer. Circuits that include large $n$-bit Toffoli gates ($n \geq 3$), such as those designed from cost-expensive gates and hard-to-decompose Exclusive-or Sum of Products (ESOP) expressions, have complications of effective mappings into contemporary quantum layouts, such as the square grid and heavy-hex layouts. These complications are primarily caused by the limited connectivity among the physical qubits in such layouts, leading to the insertion of many additional SWAP gates. This paper introduces a new quantum circuit synthesis methodology by exploring the advantage of a Positive Davio lattice (PDL) as an intermediate representation to create our proposed triangular layout and layout-aware circuits. From these circuits, we introduce and form the SWAT gate, composed of a SWAP gate followed by a 3-bit Toffoli gate. To illustrate the usefulness of our method for existing industrial quantum layouts, we also introduce cost-effective mappings of the resulting circuits onto square grid and heavy-hex layouts without additional SWAP gates. This is done with the help of the SWAT gate. Our research highlights PDLs as an efficient tool for layout-aware quantum circuit synthesis.
}{}{}

\vspace*{10pt}

\vspace*{3pt}
\communicate{to be filled by the Editorial}

\vspace*{1pt}\textlineskip

% Introduction
\section{Introduction}
Quantum computing has emerged as a powerful tool with the potential to solve problems across various fields, offering performance that surpasses that of classical computers \cite{arute}. However, to have effective quantum computers, awareness of the quantum layout, the arrangement of physical qubits in a quantum computer, is a crucial design factor for cost-effectively routing and mapping the logical qubits of a quantum circuit. The connectivity in a quantum layout determines which qubits can interact with each other. To operate on a pair of physical qubits that are not directly connected in a layout, SWAP gates are required. The number of additional SWAP gates has a significant impact on the quantum cost \cite{maslov, albayaty-gala, albayaty-cala, liu}, especially on ESOP (Exclusive-or Sum of Products) expressions composed from a set of $n$-bit Toffoli gates, where $n \geq 3$ qubits \cite{schmitt, wille}. 

ESOP expressions are widely used in designing and constructing quantum circuits. For example, for an ESOP expression $f= abc \oplus b\bar{c}d \oplus \bar{a}cde$, the input qubits are $a$, $b$, $c$, $d$, and $e$, and additional ancilla qubits may be used to realize intermediate operations of $n$-bit Toffoli gates. $n$-bit Toffoli gates of $n-1$ control qubits and 1 target qubit are mainly utilized to construct different quantum gates and operators, such as the Fredkin gate \cite{lapierre}, Miller gate \cite{yangg}, and arithmetic adders and comparators \cite{donaire}. Therefore, minimizing the number of SWAP gates in a given layout is a critical methodology for cost-effectively synthesizing an ESOP expression or any circuit with a large number of $n$-bit Toffoli gates for a real quantum computer.

Two well-known industrial quantum layouts are the square grid layouts \cite{arute, helmer, gong}, used by Google and Rigetti, and heavy-hex layouts \cite{chamberland}, used by IBM. However, circuits with a high number of $n$-bit Toffoli gates are cost-expensive \cite{maslov, albayaty-gala, albayaty-cala} to map onto both layouts and require many additional SWAP gates. 

In our previous work \cite{perkowski}, a unique and promising method of synthesizing quantum circuits with Positive Davio lattices \cite{jeske,lukac,gao}, lattice diagrams based on the \textit{Positive Davio expansion} \cite{davio}, was introduced. However, in \cite{perkowski}, there has yet to be any relation for this method to realistic contemporary layouts. We call a Positive Davio lattice a ``PDL". The Positive Davio expansion has a relation to one of the outputs of the 3-bit Toffoli gate, an important and commonly-used gate with two controls and one target, introduced in \cite{toffoli}. This circuit synthesis technique takes advantage of the regular structure of these lattices and the relationship of Positive Davio and Toffoli gates to synthesize circuits with only 3-bit Toffoli and SWAP gates. From PDL-based circuits, we introduce the ``SWAT gate," a SWAP gate followed by a 3-bit Toffoli gate, which is an effective intermediate step for mapping the logical to physical qubits of a PDL-based circuit onto a regular quantum layout.

In this paper, we extend our past research on synthesizing quantum circuits from PDLs \cite{perkowski,lukac}. We also introduce mappings of the connectivity between the qubits in these PDL-based circuits onto a new triangular layout, the square grid layout, and the heavy-hex layout, without adding additional SWAP gates that are not already contained inside a PDL-based circuit. This methodology is aimed at reducing the cost of both circuit synthesis and the implementation of these circuits onto physical qubits. Practically, we compare the quantum cost after applying the IBM Qiskit Transpiler~\cite{qiskit} on the circuits resulting from our PDL-based method, used together with the CALA-$n$~\cite{albayaty-cala} decomposition of the 3-bit Toffoli gate, with the ESOP-based method~\cite{schmitt, wille}. We conclude that our PDL-based method generally has lower quantum costs compared to the ESOP-based method. 

This paper is organized as follows. Section 2 provides background and related work on lattice-based circuit synthesis and quantum layouts. Section 3 introduces how to map PDL-based circuits onto the current square grid layout, current heavy-hex layout, and our newly introduced triangular layout. Section 4 presents the results by demonstrating multiple practical examples and comparisons utilizing our method. Lastly, Section 5 concludes this research with potential future directions.

% Background
\section{Background and Related Work}
This section covers PDLs, the functions in these lattices, and quantum layouts.

\subsection{Expansions and Positive Davio Lattices (PDLs)}
This subsection is divided into Boolean expansions \cite{davio,shannon} and Positive Davio lattices \cite{lukac,gao}.

Logic is represented in mathematical terms with Boolean algebra \cite{kohavi}. In Boolean algebra, variables can only take on the value of either 0 (false/off) or 1 (true/on). Similar to other algebras, Boolean functions are represented with equations and variables.

Some important identities~\cite{wakerly} to note are: $x \oplus x = 0$, $0 \oplus x = x$, $1 \oplus x = \bar{x}$. These are used for simplifying Boolean functions.

\subsubsection{Boolean Expansions}
Boolean functions can be expanded using expansions. One expansion is stated in Eq.~\eqref{eqn:shannon}. This is known as the \textit{Shannon expansion} \cite{shannon}. The terms $f_{\bar{a}}$ and $f_a$ are the negative and positive cofactors, respectively. The negative cofactor $f_{\bar{a}}$ is the function $f$ when the variable $a$ is substituted with 0. Similarly, the positive cofactor $f_a$ is the function $f$ when the variable $a$ is substituted with 1.

\begin{equation}
f = \bar{a}f_{\bar{a}} + af_a
\label{eqn:shannon}
\end{equation}

In Eq.~\eqref{eqn:shannon}, we can also replace the OR(+) operation with the XOR($\oplus$) operation, as stated in Eq.~\eqref{eqn:shannon-xor}, because the two terms are \textit{disjoint}, meaning that both terms cannot both equal 1 at the same time.

\begin{equation}
f = \bar{a}f_{\bar{a}} \oplus af_a
\label{eqn:shannon-xor}
\end{equation}

\noindent 
\\\textbf{Example 1:} Let's take an example with the function $f = a \oplus \bar{a}b \oplus ac \oplus c$. The negative cofactor would be $f_{\bar{a}} = (0) \oplus (1)b \oplus (0)c \oplus c = b \oplus c$. 

The positive cofactor would be $f_a = (1) \oplus (0)b \oplus (1)c \oplus c = 1$. Using the Shannon expansion formula in Eq.~\eqref{eqn:shannon-xor}, we can get the following from the positive and negative cofactors:
$$f = \bar{a}(b \oplus c) \oplus a(1) = \bar{a}b \oplus \bar{a}c \oplus a.$$

The Shannon expansion is not the only possible expansion of Boolean functions. There are also the \textit{Davio expansions} \cite{davio}. They are derived from the Shannon expansion theorem from Eq.~\eqref{eqn:shannon-xor}. 

To derive the \textit{Positive Davio expansion}, substitute $\bar{a} = 1 \oplus a$ into Eq.~\eqref{eqn:shannon-xor} to obtain $f = (1 \oplus a)f_{\bar{a}} \oplus af_a$  and simplify to Eq.~\eqref{eqn:positive-davio}.

\begin{equation}
f = f_{\bar{a}} \oplus a(f_{\bar{a}} \oplus f_a)
\label{eqn:positive-davio}
\end{equation}

\begin{figure}[!htb]
\centering
\includegraphics[width=0.3\textwidth]{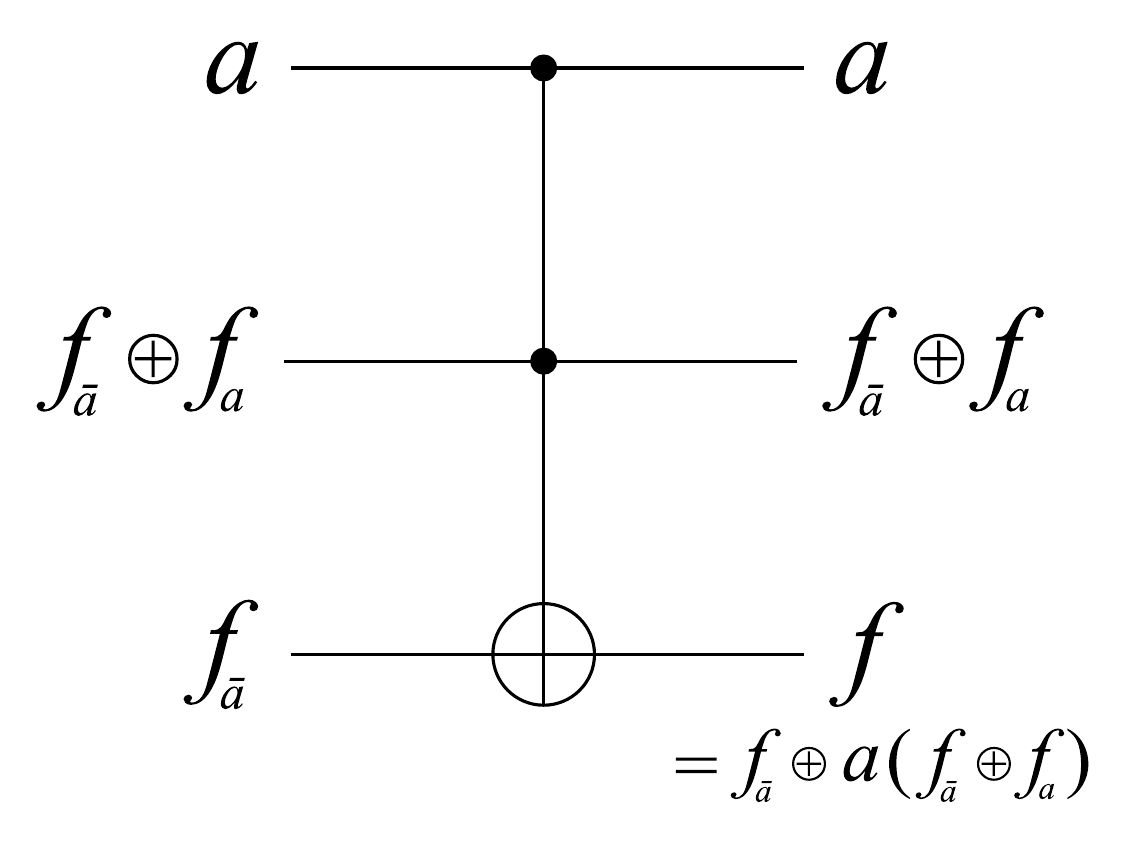}
\caption{The Positive Davio expansion realized with a Toffoli gate.}
\label{fig:fig01}
\end{figure}

The Positive Davio expansion can be tied directly to an output of the Toffoli gate, as shown in Fig.~\ref{fig:fig01}. This makes the Positive Davio expansion a valuable circuit simplification tool in quantum computing.

Similarly, we can derive the \textit{Negative Davio expansion}, stated in Eq.~\eqref{eqn:negative-davio}, by substituting $a = 1 \oplus \bar{a}$ into Eq.~\eqref{eqn:shannon-xor} to obtain $f = \bar{a}f_{\bar{a}} \oplus (1 \oplus \bar{a})f_a$ and simplifying.

\begin{equation}
f = f_a \oplus \bar{a}(f_{\bar{a}} \oplus f_a)
\label{eqn:negative-davio}
\end{equation}

The Shannon, Positive Davio, and Negative Davio expansions are valuable tools for simplifying and decomposing circuits in both classical and quantum logic.

\subsubsection{Positive Davio Lattices (PDLs)}
Lattices and trees can be formed using Boolean expansions. In the extreme cases, the number of nodes in each level of a tree increases exponentially; for lattices, this increases linearly; however, some trees grow less than exponentially. We will form the PDL \cite{lukac,gao} in this section.

\begin{figure}[!htb]
\centering
\includegraphics[width=0.25\textwidth]{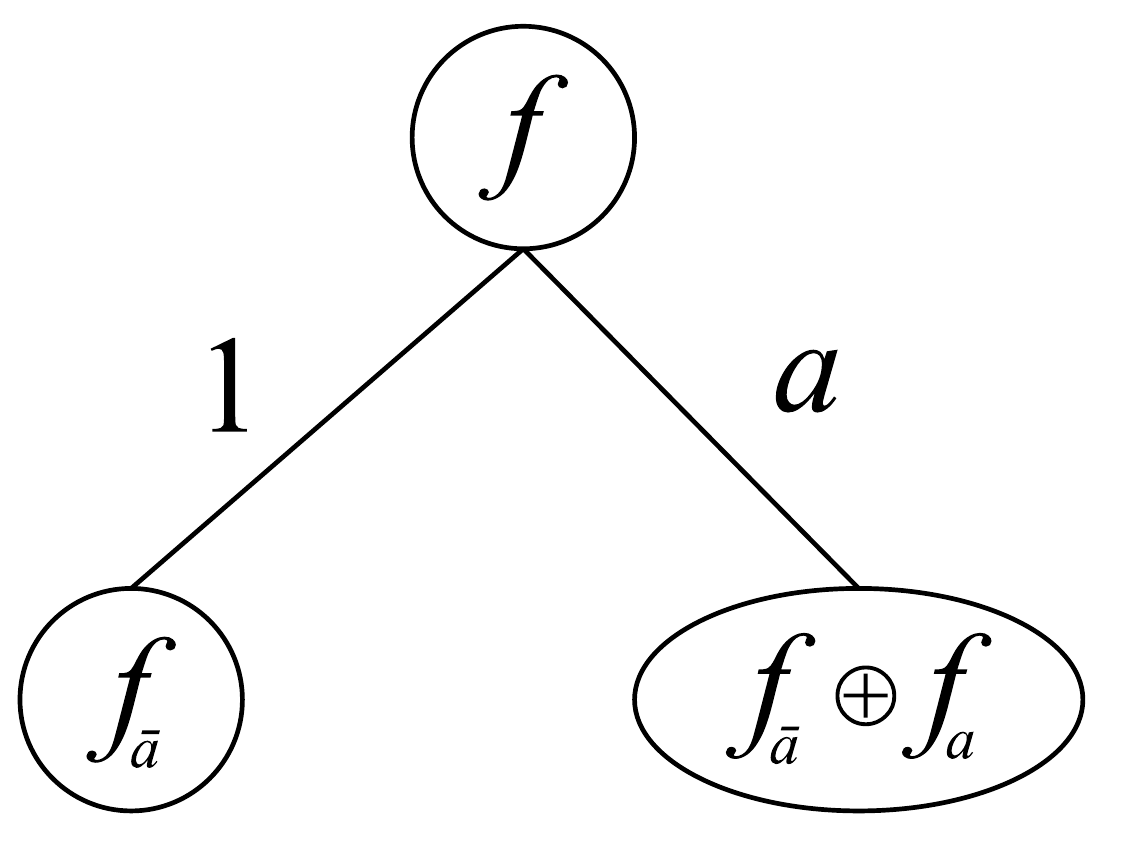}
\caption{Two levels of a PDL showing a single Positive Davio expansion: \\ $f = f_{\bar{a}} \oplus a(f_{\bar{a}} \oplus f_a)$.}
\label{fig:fig02}
\end{figure}

A level of a PDL is realized as described in Fig.~\ref{fig:fig02}. This is seen clearly in the top level of the lattice. The nodes represent Boolean functions. The edges represent the Boolean multiplication (AND operation) that occurs in a Positive Davio expansion, while the meeting of these edges represents the XOR operation.

\begin{figure}[!htb]
\centering
\includegraphics[width=0.7\textwidth]{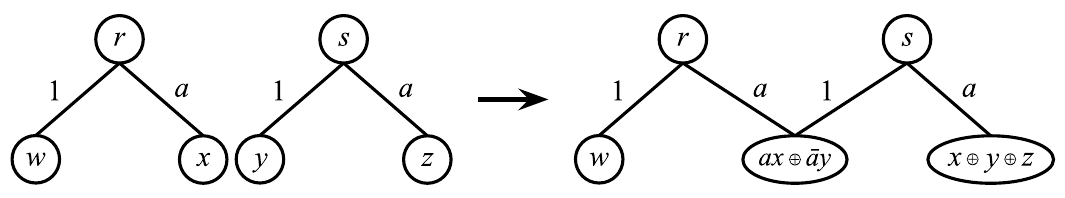}
\caption{Merging the middle nodes of two adjacent expansions using the XOR operation.}
\label{fig:fig03}
\end{figure}

The adjacent nodes of two adjacent expansions in a PDL can be merged (combined), as illustrated in Fig.~\ref{fig:fig03} \cite{lukac}. In general, the operation of such an expansion is well-known in trees~\cite{suthaharan, quinlan} and decision diagrams~\cite{drechsler}. In decision diagrams, the merging operation is used on isomorphic nodes. As introduced in~\cite{lukac}, lattices can use a third type of operation, which is the merging of arbitrary, not necessarily isomorphic, nodes. Arbitrary means arbitrary switching functions of nodes, in this case, binary Boolean functions.

This can be verified. From the two individual parts on the left of Fig.~\ref{fig:fig03}, it can be found that $r = w \oplus ax$ and $s = y \oplus az$ from the Positive Davio relation. 

The values $r$  and $s$  can also be calculated using the lattice on the right of Fig.~\ref{fig:fig03}. This yields $r = w \oplus a(ax \oplus \bar{a}y) = w \oplus ax$ and $s = (ax \oplus \bar{a}y) \oplus a(x \oplus y \oplus z) = ax \oplus \bar{a}y \oplus ax \oplus ay \oplus az = y \oplus az$, the same values as earlier.

\noindent 
\\\textbf{Example 2:} Let’s build a lattice with the function $f = 1 \oplus ad \oplus bd \oplus abd \oplus ac \oplus bc \oplus cd \oplus bcd.$ To start, we can use the Positive Davio expansion with $c$. Although we use the variable $c$  in this example first, any variable can be used. 

\begin{figure}[!htb]
\centering
\begin{subfigure}[!htb]{0.45\linewidth}
    \centering
    \includegraphics[width=\linewidth]{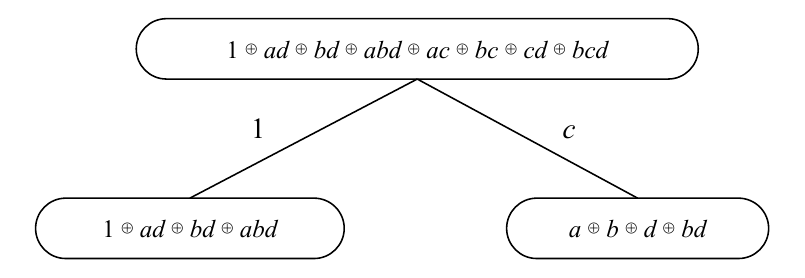}
    \caption{First two levels of a PDL for $f$.}
    \label{fig:fig04}
\end{subfigure}
\begin{subfigure}[!htb]{0.5\linewidth}
    \centering
    \includegraphics[width=\linewidth]{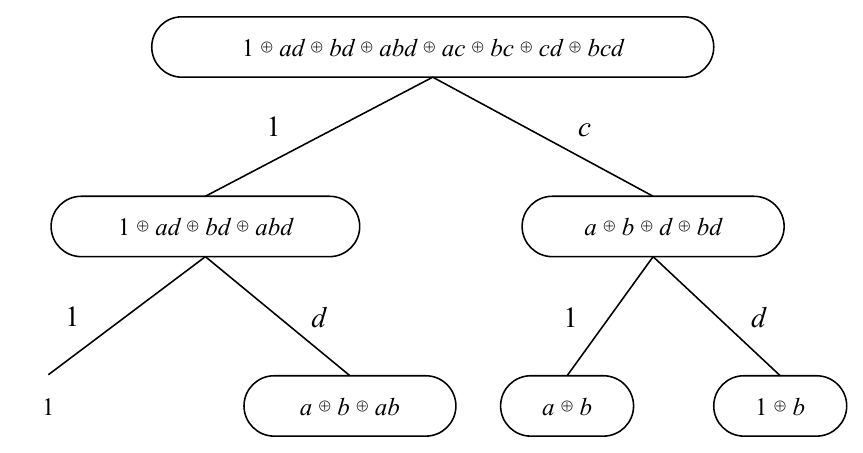}
    \caption{First three levels of a PDL for $f$.}
    \label{fig:fig05}
\end{subfigure}
\begin{subfigure}[!htb]{0.5\linewidth}
    \centering
    \includegraphics[width=\linewidth]{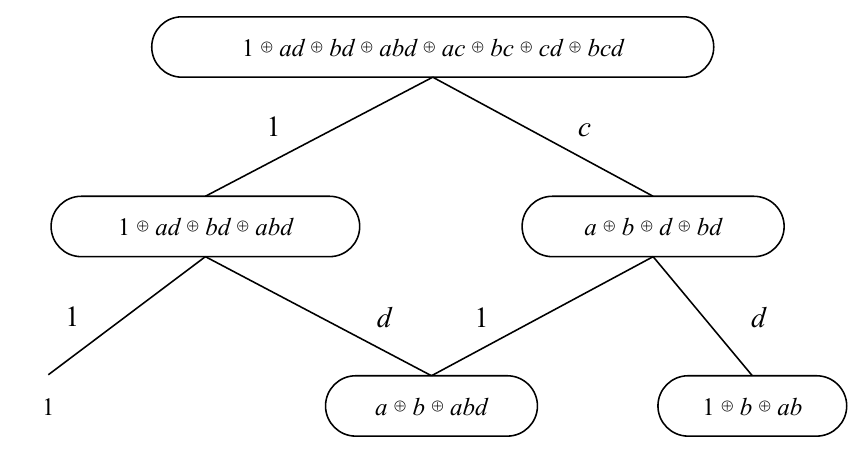}
    \caption{First three levels of a PDL for $f$ with a shared node.}
    \label{fig:fig06}
\end{subfigure}
\caption{Levels of the PDL for $f = 1 \oplus ad \oplus bd \oplus abd \oplus ac \oplus bc \oplus cd \oplus bcd.$}
\end{figure}

The negative cofactor is $f_{\bar{c}} = 1 \oplus ab \oplus bd \oplus abd$ and the positive cofactor is  $f_c = 1 \oplus ad \oplus bd \oplus abd \oplus a \oplus b \oplus d \oplus bd.$ The XOR of the cofactors is $f_{\bar{c}} \oplus f_c = a \oplus b \oplus d \oplus bd.$

Using the negative cofactor and the XOR of the cofactors, the first two levels of the lattice are built as shown in Fig.~\ref{fig:fig04}. Applying the same process to both of the nodes in the second level, the lattice can be expanded as shown in Fig.~\ref{fig:fig05}.

The two middle nodes of the third level can be merged. Merging the nodes gets $d(a \oplus b \oplus ab) \oplus \bar{d}(a \oplus b)$, which simplifies to $a \oplus b \oplus abd$ for the center node. 

For the rightmost node: $(a \oplus b \oplus ab) \oplus (a \oplus b) \oplus (1 \oplus b) = 1 \oplus b \oplus ab$. The third level can be redrawn as shown in Fig.~\ref{fig:fig06}.

\begin{figure}[!htb]
\centering
\includegraphics[width=0.5\textwidth]{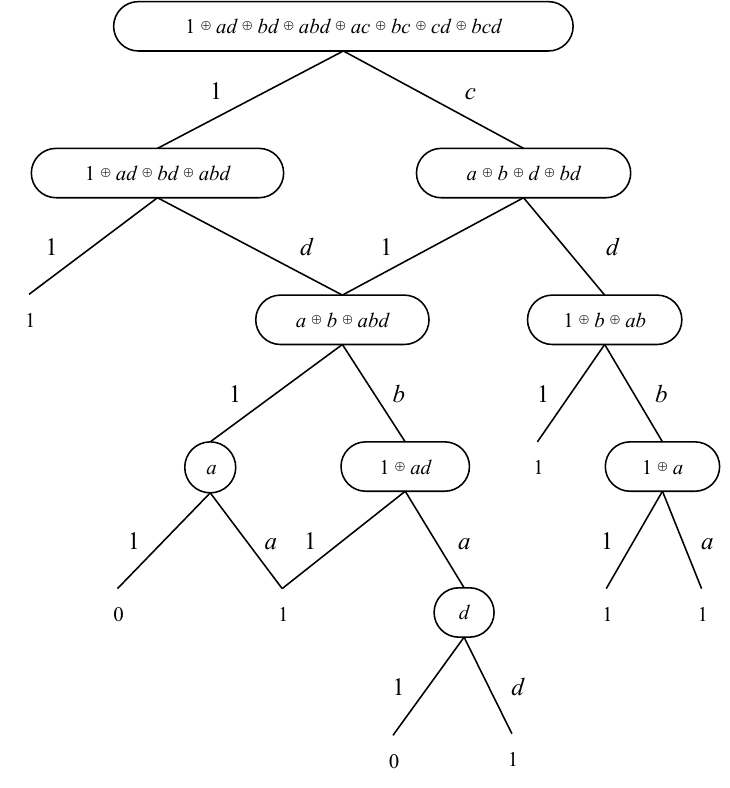}
\caption{PDL for $f = 1 \oplus ad \oplus bd \oplus abd \oplus ac \oplus bc \oplus cd \oplus bcd$.}
\label{fig:fig07}
\end{figure}

Subsequently, by repeating this process until all the nodes reach 0 or 1, the full lattice is obtained, as shown in Fig.~\ref{fig:fig07}. All nodes with child nodes that are only 0 or 1 can also be simplified, such as replacing $1 \oplus a$ with $\bar{a}$.

In the case of symmetric functions with $n$ variables, the PDL is completed at the level of $n$. Therefore, the total number of SWAP gates is $1+2+\dots+n = \frac{n(n+1)}{2}$. In the case of non-symmetric functions, some variables are repeated. The problem of which variables to repeat and in which order is still open despite several published papers. This problem is discussed in Section 2.3. The numerical results from previous research were promising, and the number of repetitions was usually not more than 3. Symmetric functions will be elaborated on in Sections 2.2.1 and 2.2.2. The optimization of the number of variable repetitions, in other words, finding the optimal variable ordering, is a difficult and open problem; however, randomized algorithms create good solutions.

\subsection{Functions in PDLs}
This subsection starts with an introduction to symmetric functions, proceeds with an explanation of their relation to PDLs, and ends with a method of synthesizing circuits from these lattices.

\subsubsection{Symmetric Functions}
A \textit{totally symmetric function} is a Boolean function where the resulting function is equivalent to its original if any two variables are swapped \cite{kohavi}. There are several other definitions of symmetric functions, but in this paper, we use a standard totally symmetric function from \cite{kohavi}. To get a greater understanding of what symmetric functions are, let’s look at the function below:
$$f = \bar{a}\bar{b}c + \bar{a}b\bar{c} + a\bar{b}\bar{c}.$$

\begin{figure}[!htb]
\centering
\begin{subfigure}[!htb]{0.15\textwidth}
    \centering
    \includegraphics[width=\textwidth]{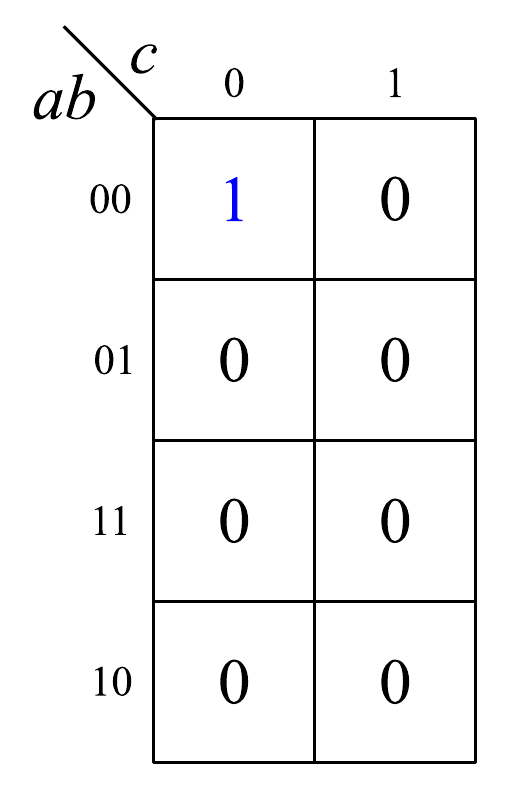}
    \caption{$\text{S}^{0}(a,b,c)$.}
\end{subfigure}
\begin{subfigure}[!htb]{0.15\textwidth}
    \centering
    \includegraphics[width=\textwidth]{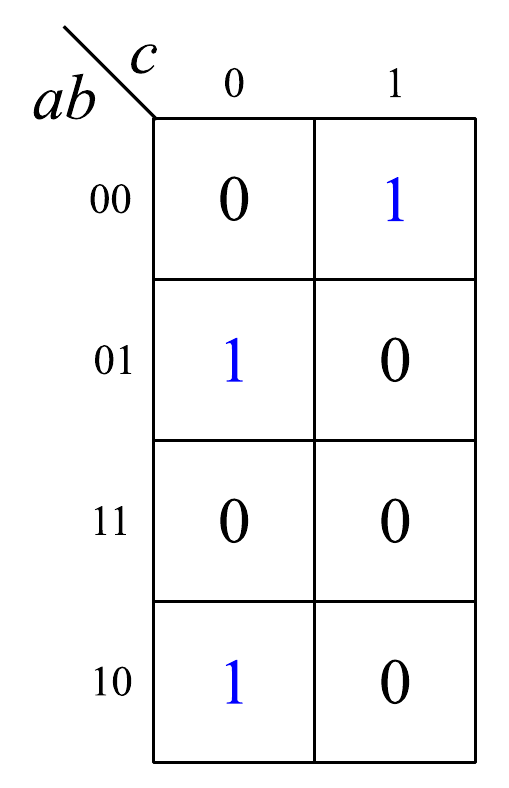}
    \caption{$\text{S}^{1}(a,b,c)$.}
    \label{fig:fig08}
\end{subfigure}
\begin{subfigure}[!htb]{0.15\textwidth}
    \centering
    \includegraphics[width=\textwidth]{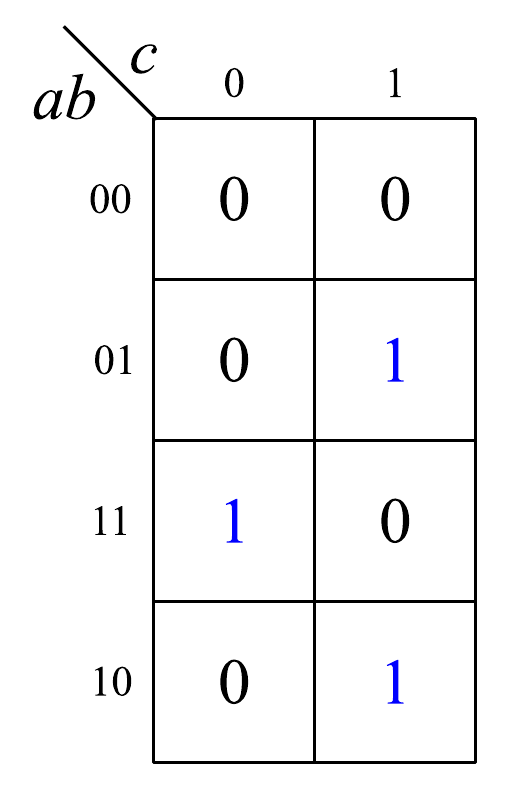}
    \caption{$\text{S}^{2}(a,b,c)$.}
\end{subfigure}
\begin{subfigure}[!htb]{0.15\textwidth}
    \centering
    \includegraphics[width=\textwidth]{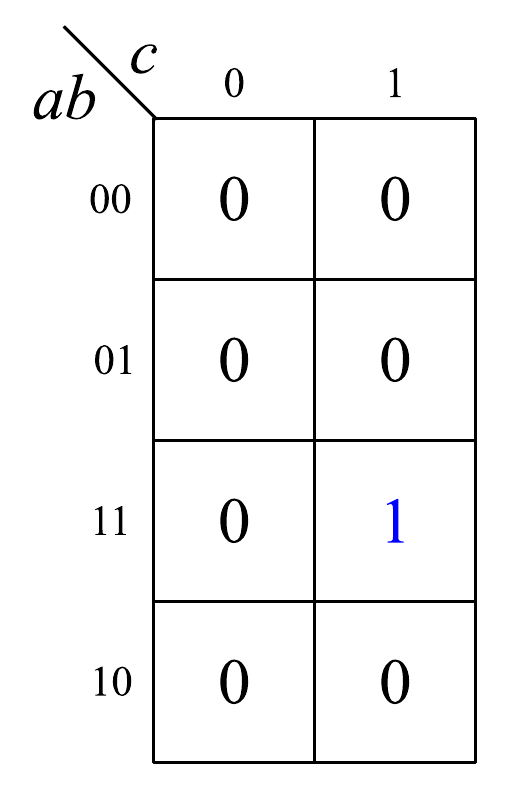}
    \caption{$\text{S}^{3}(a,b,c)$.}
\end{subfigure}
\caption{Karnaugh maps for symmetric functions.}
\label{fig:fig09}
\end{figure}

This function can be represented in a Karnaugh map \cite{karnaugh} as shown in Fig.~\ref{fig:fig08}. The rows and columns list all the possible values of the variables in an order known as Gray code. The values in each cell of the table indicate the value of the function for the corresponding input values.

This symmetric function satisfies the definition of a symmetric function that was given above, and it contains every product term with exactly 1 positive literal; a literal being a variable with either positive polarity ($x$) or negative polarity ($\bar{x}$). To denote this function, we use the notation $\text{S}^{1}(a,b,c)$, where S stands for symmetric, 1 shows that all terms with that number of positive literals are in the function, and $(a,b,c)$ is the list of all the variables.

The symmetric functions $\text{S}^{0}(a,b,c)$, $\text{S}^{1}(a,b,c)$, $\text{S}^{2}(a,b,c)$, and $\text{S}^{3}(a,b,c)$ are shown in Fig.~\ref{fig:fig09}.

\begin{figure}[!htb]
\centering
\begin{subfigure}[!htb]{0.15\linewidth}
    \centering
    \includegraphics[width=\linewidth]{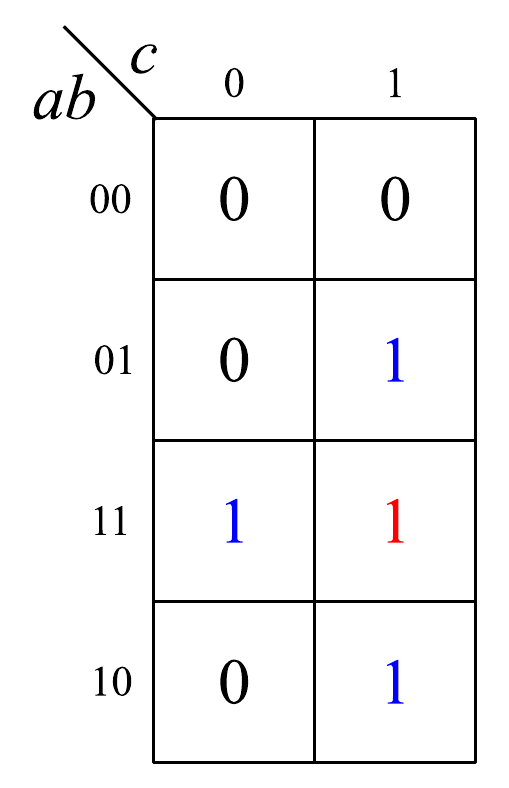}
    \caption{Karnaugh map for $\text{S}^{2,3}(a,b,c)$.}
    \label{fig:fig10}
\end{subfigure}
\begin{subfigure}[!htb]{0.3\linewidth}
    \centering
    \includegraphics[width=0.5\linewidth]{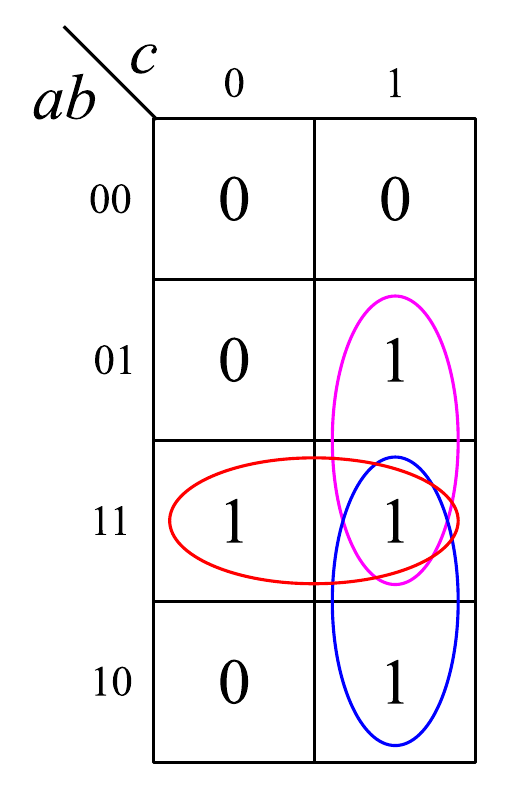}
    \caption{Karnaugh map for $\text{S}^{2,3}(a,b,c)$ with ESOP groupings.}
    \label{fig:fig11}
\end{subfigure}
\end{figure}

These aren’t the only symmetric functions. The function shown in Fig.~\ref{fig:fig10} is also symmetric. It is the sum of $\text{S}^{2}(a,b,c)$ , denoted in \textcolor{blue}{blue}, and $\text{S}^{3}(a,b,c)$ , denoted in \textcolor{red}{red}. We can write this as $\text{S}^{2,3}(a,b,c)$ . The sum of two symmetric functions is also a symmetric function.

Note that the function $\text{S}^{2,3}(a,b,c)$ can also be written as the Exclusive-OR Sum of Products (ESOP) expression $\textcolor{red}{ab} \oplus \textcolor{magenta}{bc} \oplus \textcolor{blue}{ac}$ as shown in the Karnaugh map in Fig.~\ref{fig:fig11}. The groupings (circled squares) represent the values of $f$ where the corresponding term, with the same color, in the ESOP is 1. A square that is covered by an even number of groupings is 0, while a square covered by an odd number of groupings is 1.

It can also be found that $\text{S}^{1,3}(a,b,c) = a \oplus b \oplus c$.

\subsubsection{Symmetric Functions in PDLs}
\begin{figure}[!htb]
\centering
\includegraphics[width=0.5\textwidth]{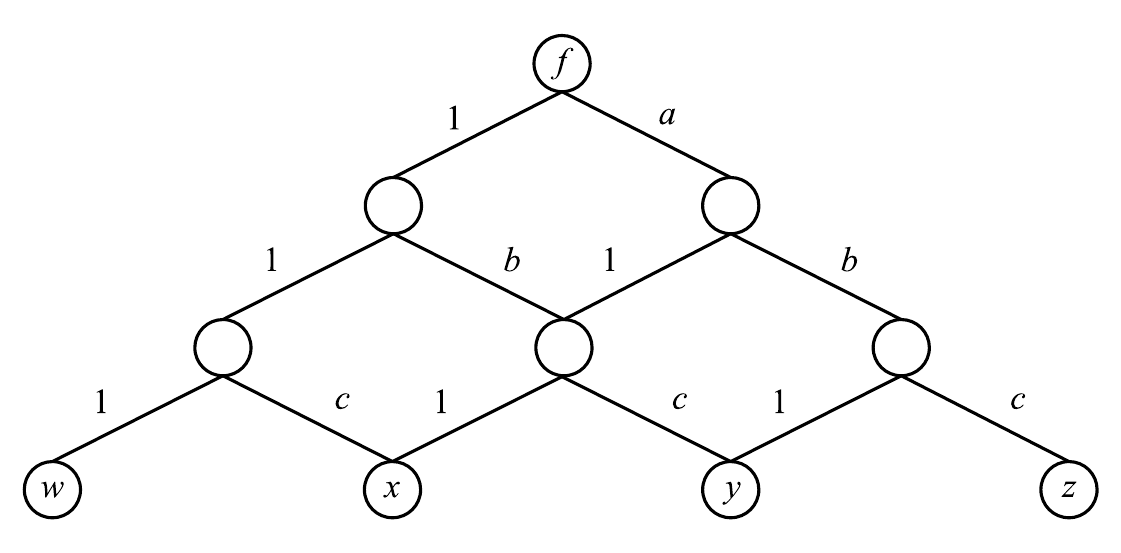}
\caption{Regular PDL.}
\label{fig:fig12}
\end{figure}

Symmetric functions can also be realized using PDLs. To do so, start with the PDL shown in Fig.~\ref{fig:fig12}.

\begin{figure}[!htb]
\centering
\includegraphics[width=0.6\textwidth]{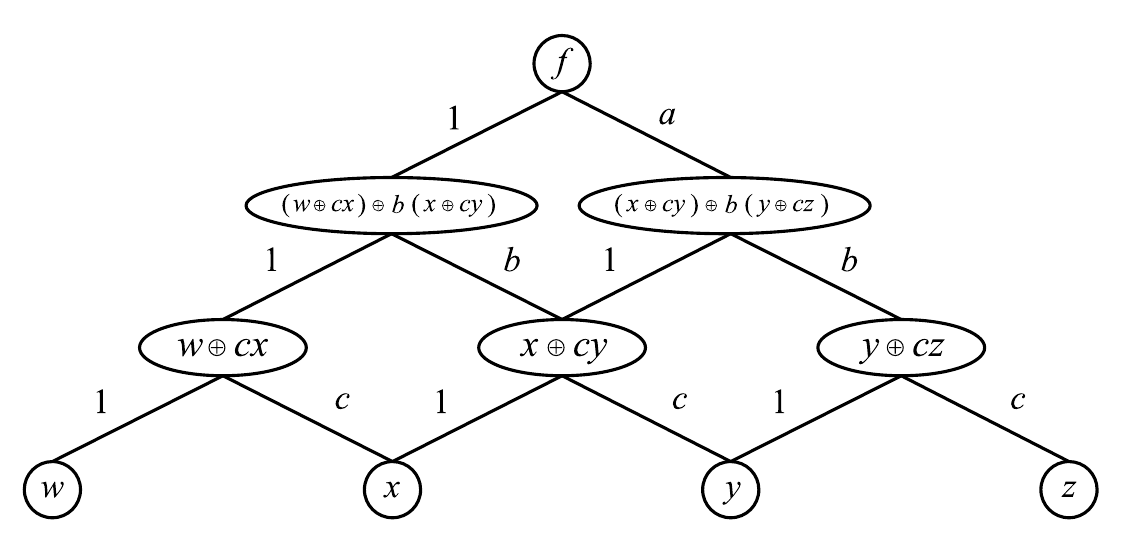}
\caption{Regular PDL with the bottom three levels filled in.}
\label{fig:fig13}
\end{figure}

Evaluating the nodes in the third level and the second level gives the values in Fig.~\ref{fig:fig13}.

Getting up to the top, it can be found that the function $f$ is:
$$f = ((w \oplus cx) \oplus b(x \oplus cy)) \oplus a((x \oplus cy) \oplus b(y \oplus cz)).$$

Expanding and simplifying, $f$ becomes:
$$f = w \oplus (a \oplus b \oplus c)x \oplus (ab \oplus bc \oplus ac)y \oplus abcz.$$

Notably, the expressions $a \oplus b \oplus c$ and $ab \oplus bc \oplus ac$ have come up earlier as symmetric functions. Therefore, $f$ can be rewritten as Eq.~\eqref{eqn:pd-sym}.

\begin{equation}
f = \text{S}^{0,1,2,3}(a,b,c)w \oplus \text{S}^{1,3}(a,b,c)x \oplus \text{S}^{2,3}(a,b,c)y \oplus \text{S}^3(a,b,c)z
\label{eqn:pd-sym}
\end{equation}

This shows PDLs can be used to realize any arbitrary symmetric function by substituting $w$, $x$, $y$, and $z$ as 0, 1, or any arbitrary function.

\begin{figure}[!htb]
\centering
\begin{subfigure}{0.5\linewidth}
    \centering
    \includegraphics[width=\linewidth]{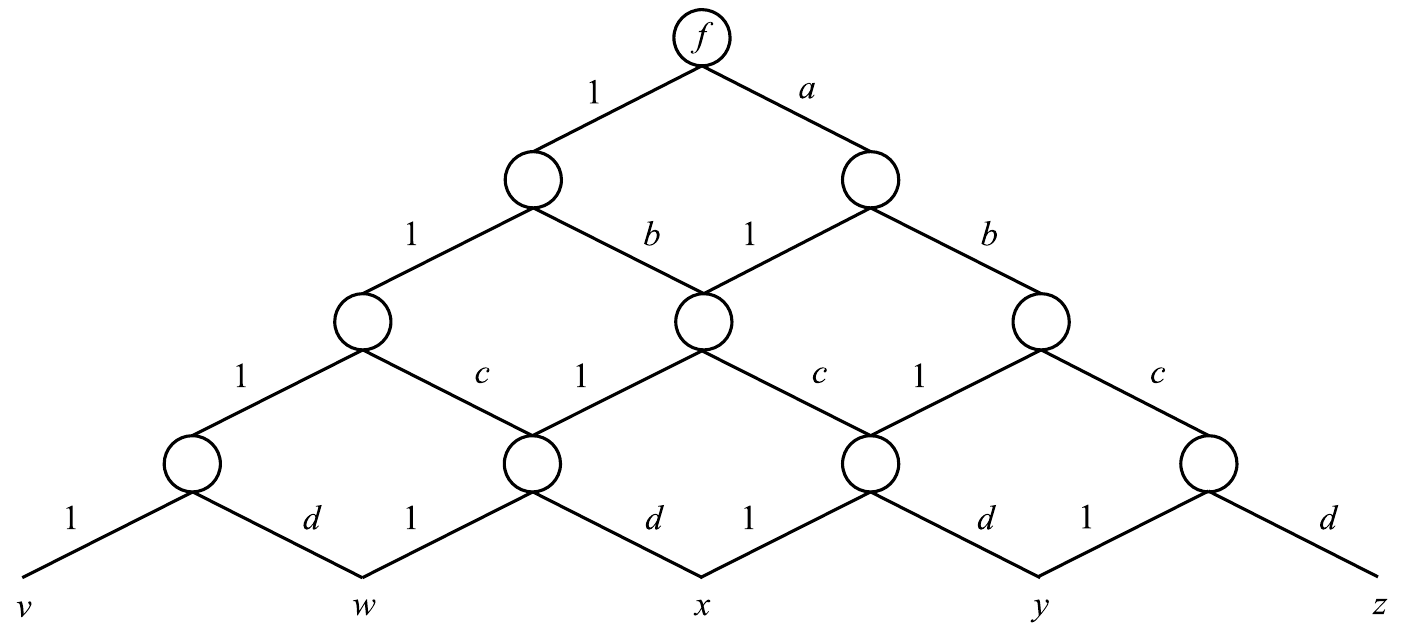}
    \caption{Regular PDL with 4 control variables.}
    \label{fig:4-var-pdl}
\end{subfigure}
\begin{subfigure}{0.24\linewidth}
    \centering
    \includegraphics[width=\linewidth]{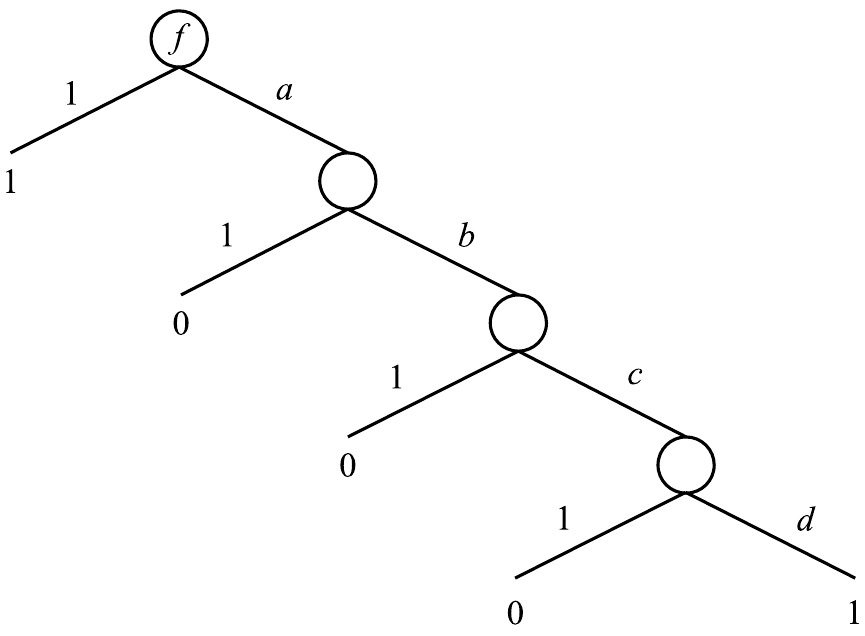}
    \caption{PDL for $\text{S}^{0,1,2,3}(a,b,c,d) = 1\oplus abcd$.}
    \label{fig:not-full-pdl}
\end{subfigure}
\begin{subfigure}{0.24\linewidth}
    \centering
    \includegraphics[width=\linewidth]{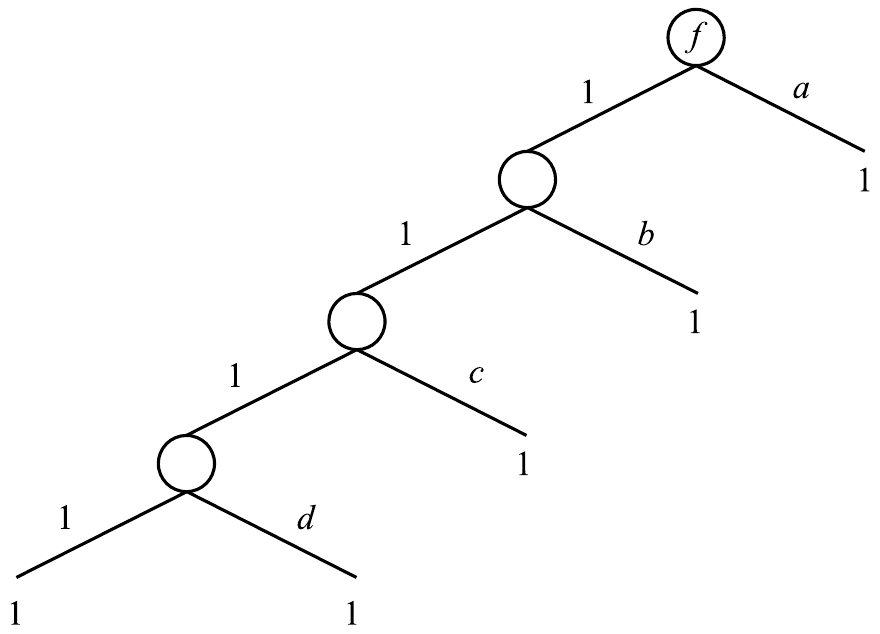}
    \caption{PDL for $\text{S}^{0,2,4}(a,b,c,d) = 1\oplus a \oplus b \oplus c \oplus d$.}
    \label{fig:affine-pdl}
\end{subfigure}
\caption{4-variable PDLs.}
\end{figure}

A regular 4-variable lattice is illustrated in Fig.~\ref{fig:4-var-pdl}. The function $f$ is equal to Eq.~\eqref{eqn:pd-sym-4-var}. Table~\ref{tab:4-var-pdl} lists all the symmetric functions along with the corresponding values for $v$, $w$, $x$, $y$, and $z$.

\begin{equation}
\label{eqn:pd-sym-4-var}
f = 
\text{S}^{0,1,2,3,4}(a,b,c,d)v \oplus 
\text{S}^{1,3}(a,b,c,d)w \oplus \text{S}^{2,3}(a,b,c,d)x \oplus \text{S}^{3}(a,b,c,d)y \oplus \text{S}^4(a,b,c,d)z
\end{equation}

\begin{table}[!htb]
    \centering
    \renewcommand{\arraystretch}{1.3}
    \resizebox{\linewidth}{!}{
    \begin{tabular}{ccccc|c||ccccc|c||ccccc|c||ccccc|c}
         $v$ & $w$ & $x$ & $y$ & $z$ & Function & 
         $v$ & $w$ & $x$ & $y$ & $z$ & Function & 
         $v$ & $w$ & $x$ & $y$ & $z$ & Function & 
         $v$ & $w$ & $x$ & $y$ & $z$ & Function \\
         \hline
         0 & 0 & 0 & 0 & 0 & $0$ &
         1 & 0 & 0 & 1 & 0 & $\text{S}^{0,1,2,4}$ &
         1 & 1 & 1 & 0 & 0 & $\text{S}^{0,3,4}$ &
         0 & 1 & 0 & 1 & 1 & $\text{S}^{1,4}$ \\

         1 & 0 & 0 & 0 & 0 & $\text{S}^{0,1,2,3,4}$ &
         1 & 0 & 0 & 0 & 1 & $\text{S}^{0,1,2,3}$ &
         1 & 1 & 0 & 1 & 0 & $\text{S}^{0,2,3,4}$ &
         0 & 0 & 1 & 1 & 1 & $\text{S}^{2,4}$ \\

         0 & 1 & 0 & 0 & 0 & $\text{S}^{1,3}$ &
         0 & 1 & 1 & 0 & 0 & $\text{S}^{1,2}$ &
         1 & 1 & 0 & 0 & 1 & $\text{S}^{0,2}$ &
         1 & 1 & 1 & 1 & 0 & $\text{S}^{0,4}$ \\

         0 & 0 & 1 & 0 & 0 & $\text{S}^{2,3}$ &
         0 & 1 & 0 & 1 & 0 & $\text{S}^{1}$ &
         1 & 0 & 1 & 1 & 0 & $\text{S}^{0,1,3,4}$ &
         1 & 1 & 1 & 0 & 1 & $\text{S}^{0,3}$ \\
         
         0 & 0 & 0 & 1 & 0 & $\text{S}^{3}$ &
         0 & 1 & 0 & 0 & 1 & $\text{S}^{1,3,4}$ &
         1 & 0 & 1 & 0 & 1 & $\text{S}^{0,1}$ &
         1 & 1 & 0 & 1 & 1 & $\text{S}^{0,2,3}$ \\

         0 & 0 & 0 & 0 & 1 & $\text{S}^{4}$ &
         0 & 0 & 1 & 1 & 0 & $\text{S}^{2}$ &
         1 & 0 & 0 & 1 & 1 & $\text{S}^{0,1,2}$ &
         1 & 0 & 1 & 1 & 1 & $\text{S}^{0,1,3}$ \\

         1 & 1 & 0 & 0 & 0 & $\text{S}^{0,2,4}$ &
         0 & 0 & 1 & 0 & 1 & $\text{S}^{2,3,4}$ &
         0 & 1 & 1 & 1 & 0 & $\text{S}^{1,2,3}$ &
         0 & 1 & 1 & 1 & 1 & $\text{S}^{1,2,3,4}$ \\

         1 & 0 & 1 & 0 & 0 & $\text{S}^{0,1,4}$ &
         0 & 0 & 0 & 1 & 1 & $\text{S}^{3,4}$ &
         0 & 1 & 1 & 0 & 1 & $\text{S}^{1,2,4}$ &
         1 & 1 & 1 & 1 & 1 & $\text{S}^{0}$ \\
    \end{tabular}}
    \caption{All 4-variable symmetric functions represented by the bottommost nodes in a PDL.}
    \label{tab:4-var-pdl}
\end{table}

Not all PDLs are completely filled in. For instance, take the lattices for the functions $\text{S}^{0,1,2,3}(a,b,c,d)$ and $\text{S}^{0,2,4}(a,b,c,d)$, shown in Fig.~\ref{fig:not-full-pdl} and Fig.~\ref{fig:affine-pdl}, which we call a ``sparse lattice," where many of the nodes are omitted since they contain a value of 0.

\subsubsection{Circuit Synthesis Using PDLs}
\begin{figure}[!htb]
\centering
\includegraphics[width=0.4\textwidth]{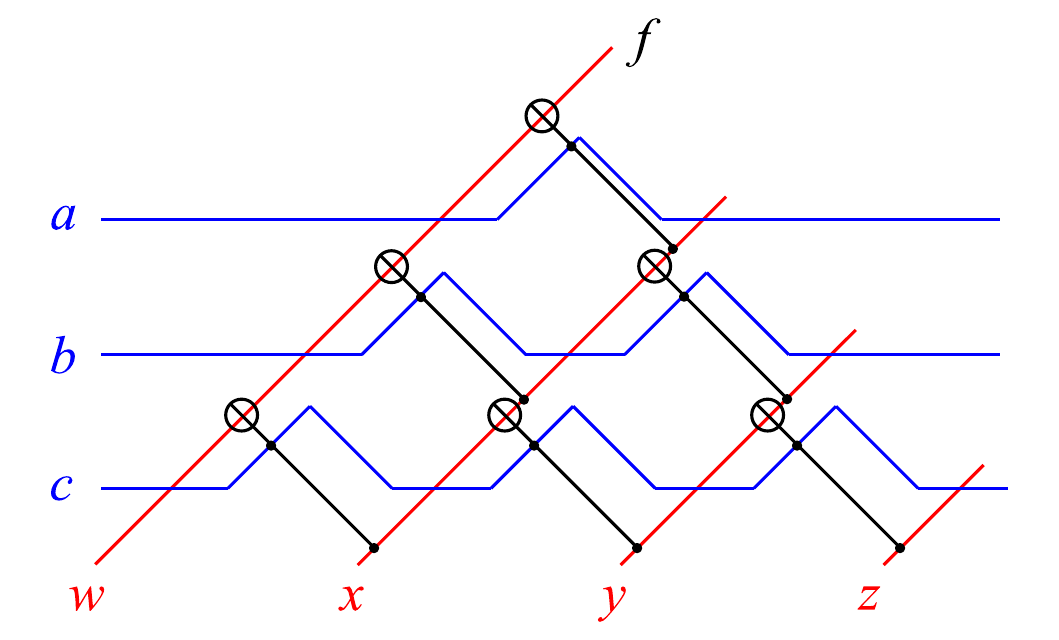}
\caption{Circuit-lattice diagram from \cite{perkowski}.}
\label{fig:fig14}
\end{figure}

Our new method of synthesizing circuits using PDLs was introduced in \cite{perkowski}. A diagram illustrating this method is in Fig.~\ref{fig:fig14}.

\begin{figure}[!htb]
\centering
\includegraphics[width=0.5\textwidth]{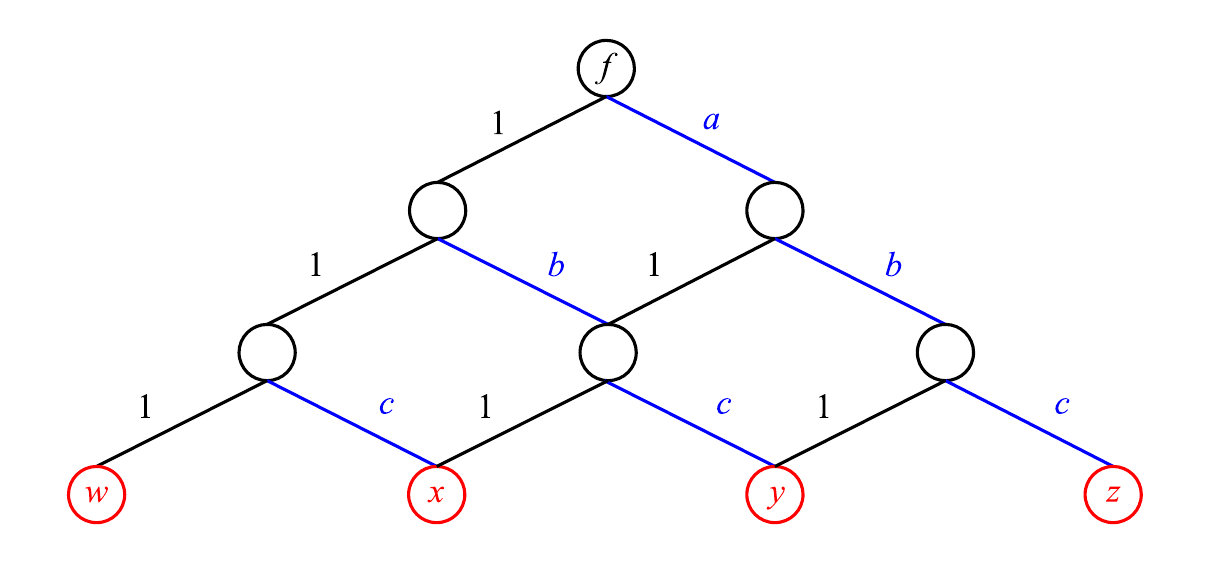}
\caption{Regular PDL.}
\label{fig:fig15}
\end{figure}

This diagram corresponds to the PDL shown in Fig.~\ref{fig:fig15}. This PDL can then be realized easily with Toffoli gates.

\begin{figure}[!htb]
\centering
\begin{subfigure}[!htb]{0.5\textwidth}
    \centering
    \includegraphics[width=\textwidth]{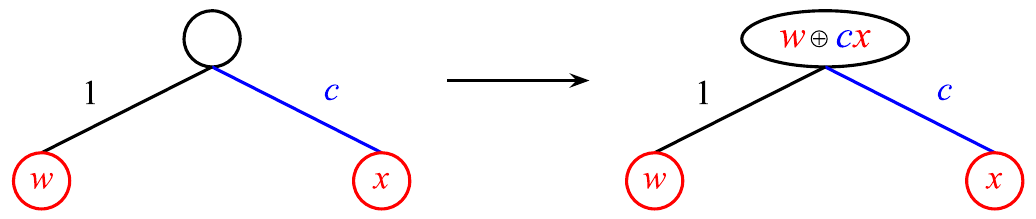}
    \caption{Part of the PDL diagram from Fig.~\ref{fig:fig15}.}
\end{subfigure}
\begin{subfigure}[!htb]{0.5\textwidth}
    \centering
    \includegraphics[width=\textwidth]{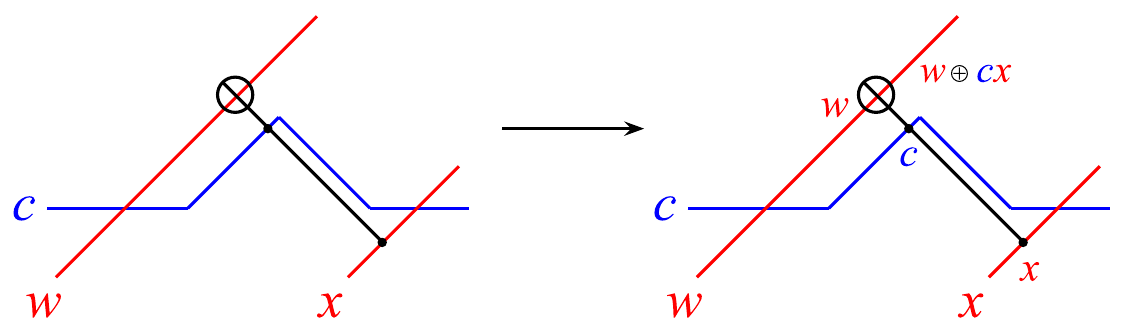}
    \caption{Part of the circuit-lattice diagram from Fig.~\ref{fig:fig14}.}
\end{subfigure}
\caption{Bottom-left part of the diagrams.}
\label{fig:fig16}
\end{figure}

Taking the bottom-leftmost part of both Fig.~\ref{fig:fig14} and Fig.~\ref{fig:fig15} gives the same expression in both diagrams, as shown in Fig.~\ref{fig:fig16}. This is true for all the parts. 

\begin{figure}[!htb]
\centering
\includegraphics[width=0.5\textwidth]{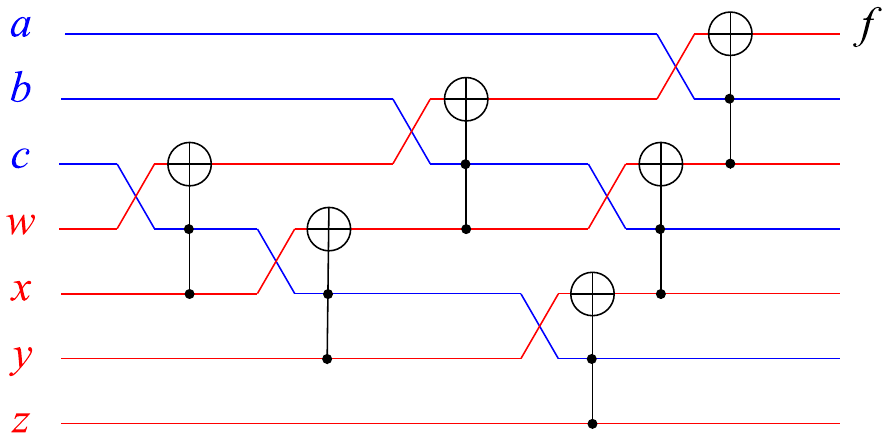}
\caption{Circuit diagram for a 7-variable function $f$ (3-level PDL) shown in standard quantum reversible gates notation.}
\label{fig:fig17}
\end{figure}

Therefore, the two diagrams correspond to the same function. Realizing the diagram as a circuit using the circuit-lattice from Fig.~\ref{fig:fig14} by twisting the lines will result in the circuit shown in Fig.~\ref{fig:fig17} with the input qubits as $a$, $b$, $c$, $w$, $x$, $y$, $z$.

\subsubsection{Circuit Synthesis for Arbitrary Functions Using PDLs}
The function $f$ that was synthesized using its PDL in the last subsection was a symmetric one. Most functions are not symmetric functions. However, our method from \cite{perkowski} can still be used to synthesize such circuits for these functions. The process of PDL-based circuit synthesis is described in Example 3.

\noindent
\\\textbf{Example 3:} The goal is to synthesize a circuit for the function $f = ab \oplus bc \oplus a\bar{c}$.

\begin{figure}[!htb]
\centering
\includegraphics[width=0.5\textwidth]{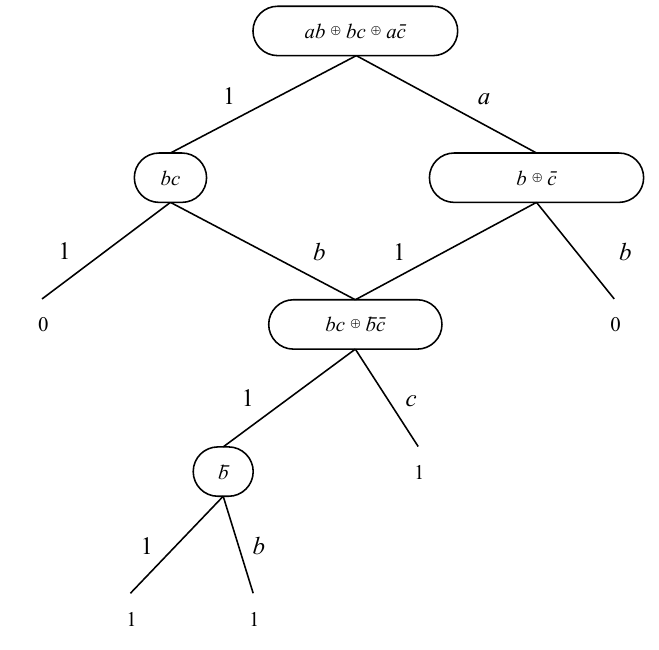}
\caption{PDL for $f = ab \oplus bc \oplus a\bar{c}$.}
\label{fig:fig18}
\end{figure}

\begin{figure}[!htb]
\centering
\begin{subfigure}{0.45\linewidth}
    \centering
    \includegraphics[width=\linewidth]{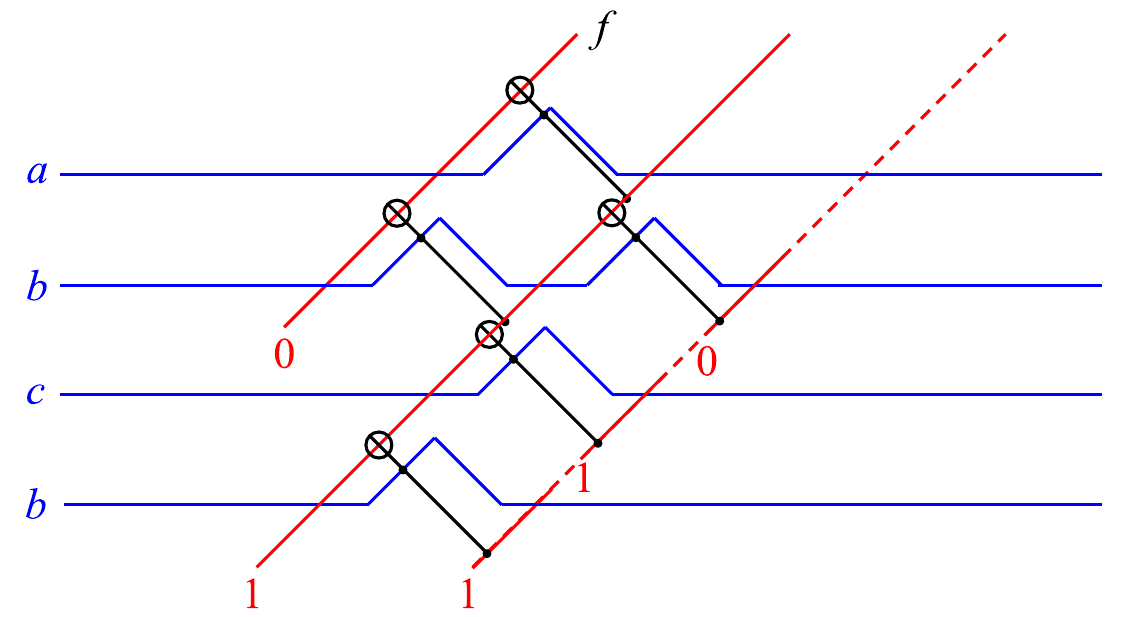}
    \caption{Circuit-lattice.}
    \label{fig:fig19}
\end{subfigure}
\begin{subfigure}{0.45\linewidth}
    \centering
    \includegraphics[width=\linewidth]{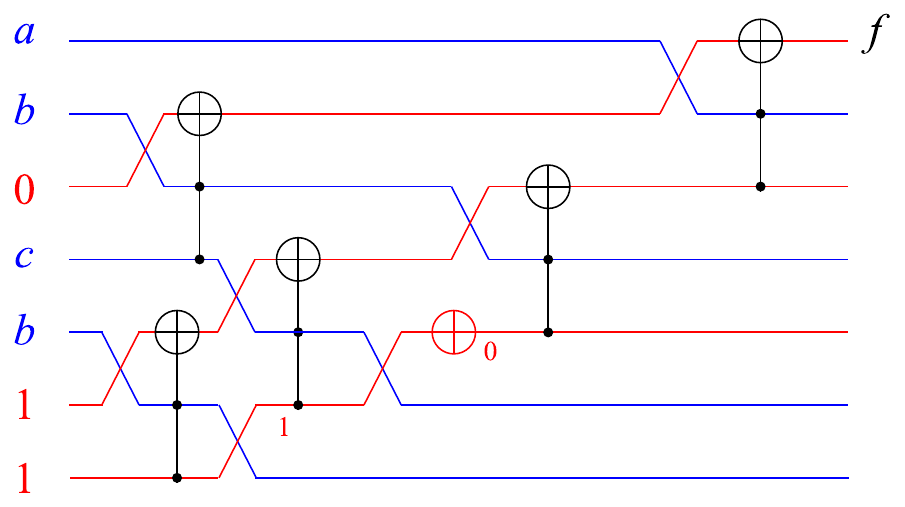}
    \caption{Circuit.}
    \label{fig:fig20}
\end{subfigure}
\caption{Circuit-lattice diagram and circuit for $f = ab \oplus bc \oplus a\bar{c}$.}
\end{figure}

\begin{figure}[!htb]

\end{figure}

The first step is to create the PDL for this function, as shown in Fig.~\ref{fig:fig18}. The next step is to create a circuit-lattice diagram from the PDL, as shown in Fig.~\ref{fig:fig19}. Lastly, from the circuit-lattice diagram, the circuit can be created, as shown in Fig.~\ref{fig:fig20}.

\subsection{Circuit Synthesis Using Other Lattices and Diagrams}
Circuits can also be synthesized from \textit{Negative Davio lattices}~\cite{jeske} (NDLs) and \textit{Shannon lattices}~\cite{jeske} by extending the method for PDLs.

\begin{figure}[!htb]
    \centering
    \begin{subfigure}{0.3\linewidth}
        \centering
        \includegraphics[width=\linewidth]{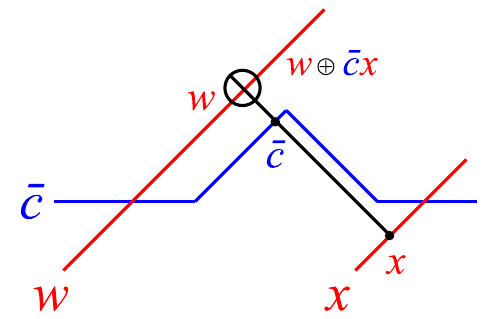}
        \caption{NDL-based.}
        \label{fig:ndl-circuit-lattice}
    \end{subfigure}
    \begin{subfigure}{0.3\linewidth}
        \centering
        \includegraphics[width=\linewidth]{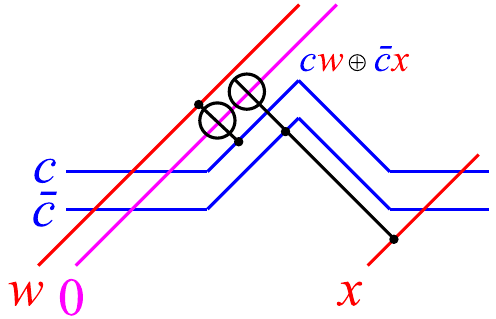}
        \caption{Shannon lattice-based.}
        \label{fig:shannon-circuit-lattice}
    \end{subfigure}
    \caption{Part of the circuit-lattice diagrams for NDL-based circuit and Shannon lattice-based circuit.}
\end{figure}

NDL-based circuits are similar to PDL-based circuits, the only difference being that instead of qubits for $a$, $b$, and $c$, there are qubits for $\bar{a}$, $\bar{b}$, and $\bar{c}$, so there are NOT gates inserted at the start of the circuit. A part of the circuit-lattice diagram for an NDL is shown in Fig.~\ref{fig:ndl-circuit-lattice}.

A part of the circuit-lattice diagram for Shannon lattice-based circuits is shown in Fig.~\ref{fig:shannon-circuit-lattice}. They are more costly, since the two products lead to two 3-bit Toffoli gates and one more qubit than what is necessary for PDL-based or NDL-based circuits. Comparing Shannon lattices and binary decision diagrams (BDDs) \cite{wille-bdd}, as both are based on only the Shannon expansion, we appreciate that they have, in general, fewer nodes. However, the nodes are more cost-expensive and require more qubits. 

There are strong relations between Davio lattices and decision diagrams, and theoretical methods based on symmetries, partial symmetries, and generalized symmetries of Boolean functions. Edwards et al. \cite{a8} created a general synthesis methodology for binary circuits based on remapping to two-variable symmetries. This method can be used in relation to lattices, but their work is not related to the complexity of lattices and the number of
introduced SWAP gates.

We, in \cite{a10}, introduced layout-aware lattices for classical technologies, including new types of generalized lattices. These lattices use Shannon, Positive Davio, and Negative Davio expansions, together with related joining operations and rotations of gates, as well as some special cases of simplified nodes. We, in \cite{jeske}, also introduced Kronecker lattices for multi-output functions that use Shannon and Davio expansions and merging techniques. Numerical results demonstrate that for several benchmarks, these designs are competitive with standard synthesis for classical logic. For instance, Drucker et al. \cite{a12} presented generalized pseudo-Kronecker symmetries and synthesis to classical lattices for generalized symmetric functions. Tabular comparison on 54 binary benchmarks demonstrates efficient results for several cases.

Work on variants of lattices was also done at Portland State University by a group of Malgorzata Chrzanowska-Jeske under the name PSBDD \cite{a13, a15, a16, a20}. Wang \cite{a13} discussed variable-ordering problems in lattices. Many experimental results on benchmarks are given in \cite{a20}. Generalized symmetries and related numerical results for various lattices and their generalizations were further discussed in \cite{a18, a19}. While only some results were directed at lattices with only positive Davio gates, these results can be extended to our lattices and the new quantum layout that are introduced here.

Extensions of lattices were also investigated at the University of California, Santa Barbara, in a group of Malgorzata Marek-Sadowska under the name of YADD \cite{a21,a22}. Good experimental results were reported, but again, the lattices with only positive Davio were not emphasized, and no relations to quantum layout were investigated.

Narahari Ramineni \cite{a14} discussed results on many benchmarks for mapping
various types of trees based on Shannon and Davio expansions. Gao et al. \cite{a9} presented various applications of symmetric functions as parts of lattice-based oracles for quantum search algorithms. These results can be used to generalize lattices to new structures that will be hybrids of lattices and small trees.

In addition to working on lattices, decision diagrams, and trees, some work in classical digital design was done on planar diagrams. Sasao, Butler, and collaborators \cite{a1,a2} discussed planar circuits that have some relation to lattices introduced in our current paper and especially to their symmetric and generalized symmetric variants. However, their research was related to BDDs, and therefore, it is based on the Shannon expansion, which is not a topic of our current methods based only on Davio expansions. Our lattices, both those that are based on positive Davio and on both Davio expansions, can be extended to planar graphs ala Sasao, Butler, and their collaborators.

In the interesting papers by Levin et al. \cite{a4} and Karpovsky et al. \cite{a3}, the authors introduced planar BDDs built using Walsh transforms, linearization, and decomposition, but their work was also related only to Shannon and not Davio expansions. All these works use decision diagrams based on expansions, and not lattice diagrams based on expansions and mergings as introduced and used here. These authors commented on an efficient realization of symmetric and other functions using planar diagrams, some of which were related to Shannon lattices. The relations of these diagrams to Davio lattices should be investigated. Davio lattices, similar to Shannon lattices, are good for symmetric functions, but the location of gates inside the lattice is very different, as shown with the sparse PDL for the function $\text{S}^{0,1,2,3}(a,b,c,d)$ in Section 2.2.2. Therefore, relations of the quantum layout to Shannon and Kronecker lattices versus Davio lattices should be discussed in the future as separate research topics.

Several other authors discussed the use of the Davio expansions in classical design. Saleem Taha \cite{a5} presented planar realizations of some Davio Decision Diagrams mapped to reversible logic, but his work was not related to the layout complexity of such realizations. Donald Chai \cite{a6} discussed circuit symmetries in classical logic synthesis and their relations to two-dimensional layout, but his work was not related to lattices. Jayati J. Law \cite{a7} presented the minimization of ancilla qubits in Ordered Kronecker Decision Diagrams (OKDDs), Ordered BDDs (OBBDs), and Ordered Functional Decision Diagrams (OFDDs). Although he discussed mappings for reversible circuits, his work was not related to lattices as well.

The tradeoff between selected types of nodes and quantum cost is an open research area, and this paper does not investigate deeper into Shannon lattice-based circuit synthesis, as this is a different problem. This paper mainly focuses on PDL-based circuits, but everything can be generalized to NDL-based circuits and circuits based on lattices with both the Positive and Negative Davio expansions, as conversion only involves inserting NOT gates. 

\subsection{Quantum Layouts}
In order to build quantum computers that are scalable and have low error rates, an effective physical quantum layout is crucial. The (physical) quantum layout, the arrangement of the qubits in a quantum computer, determines the connectivity between qubits. 

Currently, one of the most popular methods of implementing qubits is through superconduction \cite{lapierre}. The quantum layouts in these computers are generally grids with nearest-neighbor connections. In quantum computing, two of the most popular layouts are the square grid \cite{arute, helmer, gong} and heavy-hex layouts \cite{chamberland}.

\begin{figure}[!htb]
\centering
\includegraphics[width=0.2\textwidth]{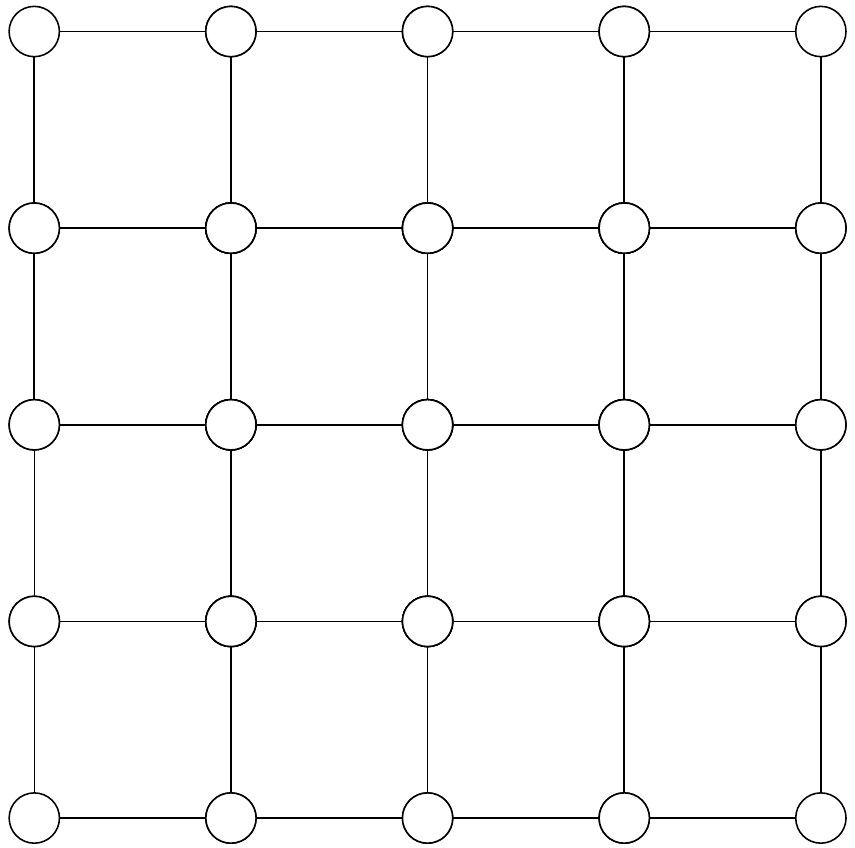}
\caption{The square grid layout.}
\label{fig:fig21}
\end{figure}

For the square grid, each qubit in this grid is connected to 4 other qubits. It was notably used in Google’s Sycamore quantum computer \cite{arute} and by Rigetti \cite{sete}, shown in Fig.~\ref{fig:fig21}. This layout is simple, scalable, and has good connectivity \cite{rached}. However, this layout also requires many SWAP gates for interactions between physical qubits that are not directly connected.

\begin{figure}[!htb]
\centering
\includegraphics[width=0.2\textwidth]{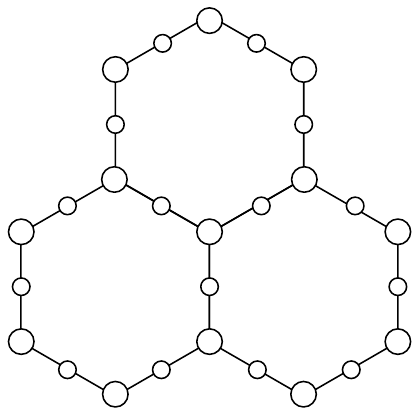}
\caption{Heavy-hex layout.}
\label{fig:fig22}
\end{figure}

For the heavy-hex layout, which is used by IBM and was introduced in \cite{chamberland}, each qubit in this grid is connected to 2 or 3 other qubits, as shown in Fig.~\ref{fig:fig22}. The benefits of this layout are its low error rate, low crosstalk, and scalability. 

These two layouts are some of the most popular layouts used by companies and academic institutions today. However, circuits that use a high number of $n$-bit Toffoli gates ($n>3$), such as circuits from ESOP expressions, are inefficient and cost-expensive for these layouts, leading to many additional SWAP gates.

\section{Mapping of Positive Davio Lattice-Based Circuits}

\subsection{Toffoli Gate Mapping and Decomposition}
The $n$-bit Toffoli gate is commonly used in constructing quantum circuits, yet implementing this gate can turn costly due to additional SWAP gates. This problem also occurs in gates that include a Toffoli, such as the Fredkin gate \cite{lapierre}. 

\begin{figure}[!htb]
\centering
\includegraphics[width=0.6\textwidth]{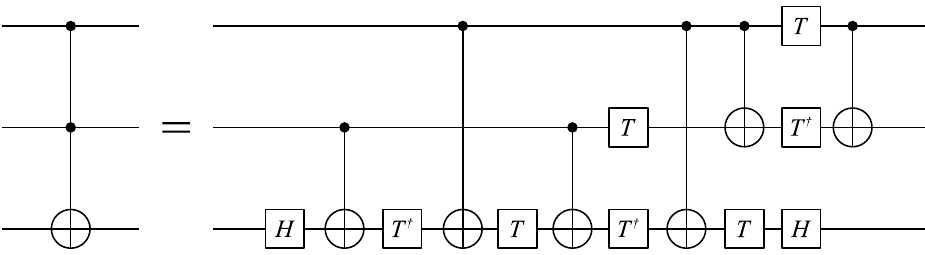}
\caption{A 3-bit Toffoli gate decomposition using the IBM Qiskit Transpiler \cite{albayaty-cala, qiskit}}
\label{fig:fig24}
\end{figure}

\begin{figure}[!htb]
\centering
\includegraphics[width=0.4\textwidth]{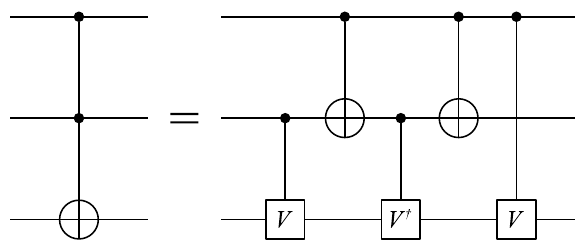}
\caption{A 3-bit Toffoli gate decomposition by Barenco et al. \cite{barenco}. The $V$ gate is the Square-Root-of-NOT gate, and the $V^\dagger$ gate is the inverse of the $V$ gate.}
\label{fig:fig25}
\end{figure}

\begin{figure}[!htb]
\centering
\includegraphics[width=0.5\textwidth]{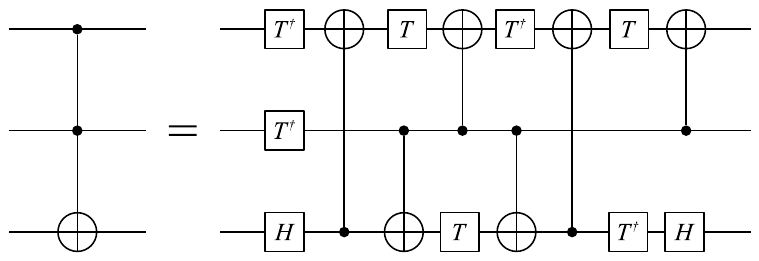}
\caption{A 3-bit Toffoli gate decomposition by Amy et al. \cite{amy}.}
\label{fig:fig26}
\end{figure}

For a 3-bit Toffoli gate to be implemented in a physical quantum computer, it must be decomposed into single-qubit and two-qubit gates. However, in many traditional 3-bit Toffoli decompositions, each of the three qubits has operations with the others. Some classical and common decompositions are illustrated in Fig.~\ref{fig:fig24} \cite{albayaty-cala, qiskit}, Fig.~\ref{fig:fig25} \cite{barenco}, and Fig.~\ref{fig:fig26} \cite{amy}.

\begin{figure}[!htb]
\centering
\begin{subfigure}[!htb]{0.3\textwidth}
    \centering
    \includegraphics[width=\textwidth]{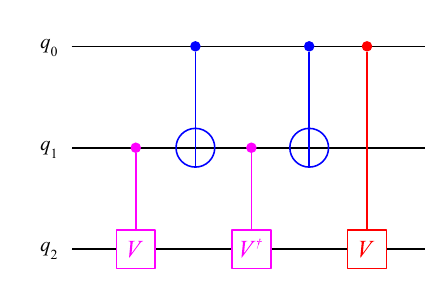}
\end{subfigure}
\begin{subfigure}[!htb]{0.2\textwidth}
    \centering
    \includegraphics[width=\textwidth]{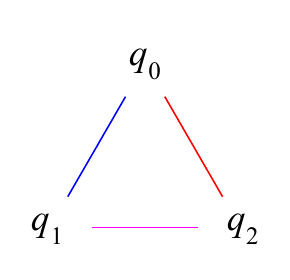}
\end{subfigure}
\caption{The triangular connection of qubits in the Toffoli gate decomposition from \cite{barenco}.}
\label{fig:fig27}
\end{figure}

To be able to realize these decompositions without additional SWAP gates, all three qubits must be connected in a triangle. An example of this is with the Barenco et al. \cite{barenco} decomposition, where the connections between qubits are illustrated in Fig.~\ref{fig:fig27}. However, in quantum layouts such as the square grid and heavy-hex, there are no triangular connections.

\begin{figure}[!htb]
\centering
\includegraphics[width=0.5\textwidth]{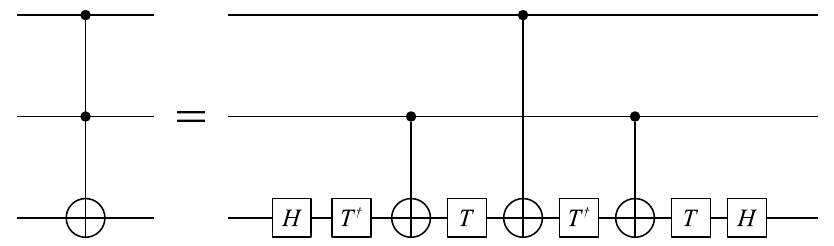}
\caption{A 3-bit Toffoli gate decomposition from \cite{albayaty-gala,albayaty-cala}.}
\label{fig:fig28}
\end{figure}

\begin{figure}[!htb]
\centering
\begin{subfigure}[!htb]{0.5\textwidth}
    \centering
    \includegraphics[width=\textwidth]{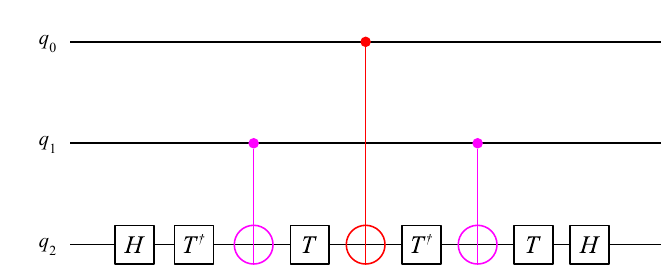}
\end{subfigure}
\begin{subfigure}[!htb]{0.2\textwidth}
    \centering
    \includegraphics[width=\textwidth]{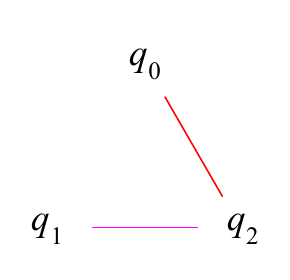}
\end{subfigure}
\caption{The non-triangular connection of qubits in the Toffoli gate decomposition from \cite{albayaty-gala,albayaty-cala}.}
\label{fig:fig29}
\end{figure}

A method of decomposing a Toffoli gate, shown in Fig.~\ref{fig:fig28} and Fig.~\ref{fig:fig29}, without using a connection between the two control qubits was introduced in \cite{albayaty-gala,albayaty-cala} (the CALA-$n$ decomposition), and it is the Qiskit ecosystem. This removes a connection between a pair of qubits. Notice that, in Fig.~\ref{fig:fig28}, all utilized gates are Clifford+T native gates, which are supported by many quantum computers \cite{albayaty-cala}.

From this decomposition, we can find the connectivity between all qubits in a PDL-based circuit.

\subsection{Triangular Mapping}
\begin{figure}[!htb]
\centering
\begin{subfigure}[!htb]{0.4\textwidth}
    \centering
    \includegraphics[width=\textwidth]{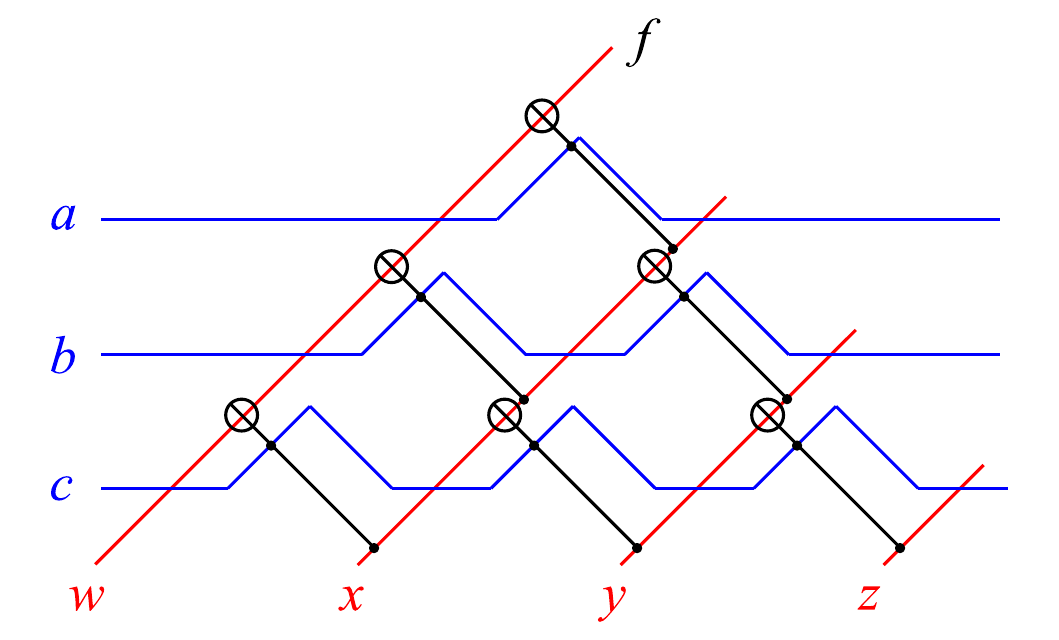}
    \caption{Circuit-lattice diagram.}
\end{subfigure}
\begin{subfigure}[!htb]{0.4\textwidth}
    \centering
    \includegraphics[width=\textwidth]{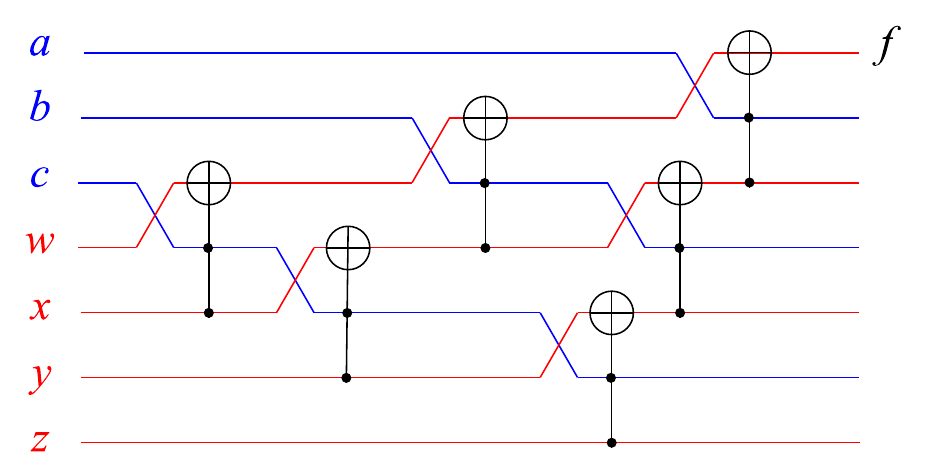}
    \caption{Quantum circuit diagram.}
\end{subfigure}
\caption{Diagrams of circuits using a PDL \cite{perkowski}.}
\label{fig:fig30}
\end{figure}

A triangular layout is tied to the PDL-based circuit synthesis method described in Sections 2.2.3 and 2.2.4, as shown in Fig.~\ref{fig:fig30} \cite{perkowski}.

\begin{figure}[!htb]
\centering
\includegraphics[width=0.2\textwidth]{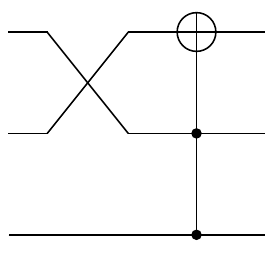}
\caption{The SWAT gate consists of one SWAP gate followed by one 3-bit Toffoli gate.}
\label{fig:fig31}
\end{figure}

This method creates circuits that use many blocks of a SWAP gate, between the top two qubits, followed by a 3-bit Toffoli, with the target as the top qubit. This combination of the SWAP gate followed by a Toffoli gate, we term the ``SWAT gate,” as presented in Fig.~\ref{fig:fig31}. The SWAT gate is an effective building block for PDL-based circuits, plus can be decomposed and used for mapping onto regular quantum layouts.

\begin{figure}[!htb]
\centering
\begin{subfigure}[!htb]{0.4\textwidth}
    \centering
    \includegraphics[width=\textwidth]{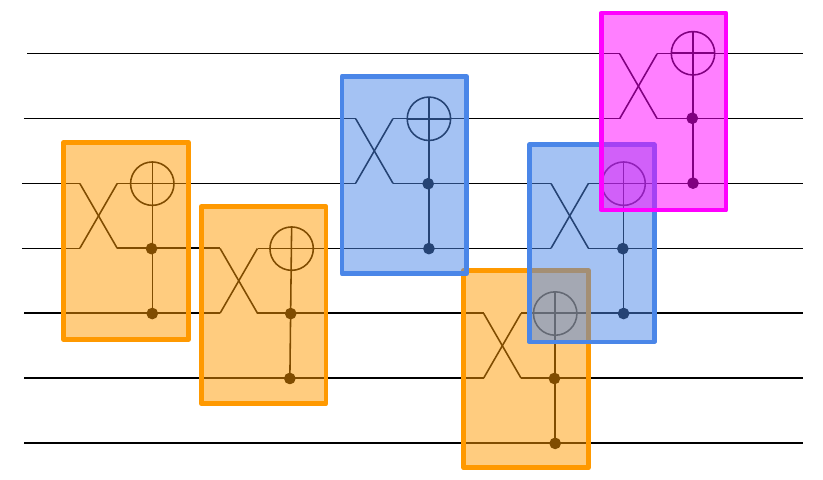}
    \caption{SWAT gates in the circuit diagram.}
\end{subfigure}
\begin{subfigure}[!htb]{0.4\textwidth}
    \centering
    \includegraphics[width=\textwidth]{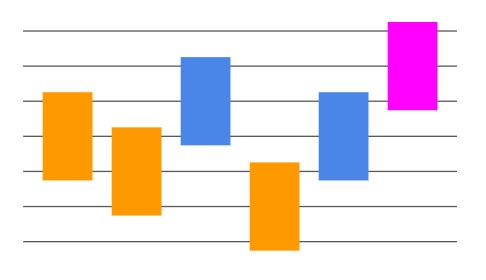}
    \caption{The circuit where the SWAT gates are represented by blocks.}
\end{subfigure}
\caption{SWAT gates in the PDL-based circuit. Different colors represent the different levels of the PDL.}
\label{fig:fig32}
\end{figure}

The circuit synthesized from the PDL is composed of a regular structure with many SWAT gates, as illustrated in Fig.~\ref{fig:fig32}.

To avoid confusion, we introduce the following terms: \textit{local} and \textit{global} SWAP gates. A local SWAP gate is a SWAP gate contained inside a SWAT gate. A global SWAP gate is a SWAP gate outside of any SWAT gate. For example, additional SWAP gates that would be added due to the limited connectivity of a layout are global.

To map the SWAT gate onto a physical quantum layout, the SWAT gate is required to be decomposed into native single-qubit and two-qubit gates supported by a quantum computer.

\begin{figure}[!htb]
\centering
\includegraphics[width=0.35\textwidth]{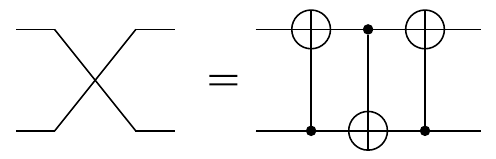}
\caption{Decomposition of one SWAP gate into three CNOT gates.}
\label{fig:fig33}
\end{figure}

In general, the SWAP gate can be decomposed into three CNOTs, as presented in Fig.~\ref{fig:fig33}.

\begin{figure}[!htb]
\centering
\includegraphics[width=0.5\textwidth]{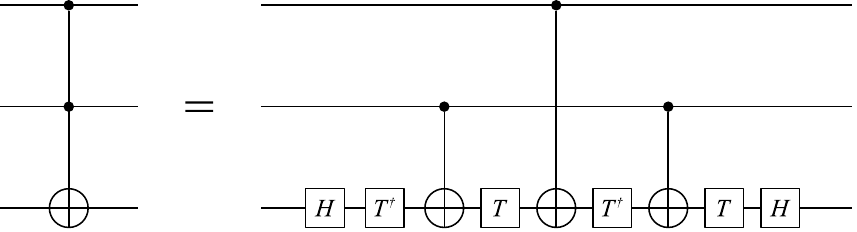}
\caption{Decomposition of the SWAT gate.}
\label{fig:fig34}
\end{figure}

Using the SWAP gate decomposition from Fig.~\ref{fig:fig33} and the Toffoli gate decomposition from \cite{albayaty-gala,albayaty-cala} in Fig.~\ref{fig:fig28}, the SWAT gate can be decomposed as shown in Fig.~\ref{fig:fig34}.

\begin{figure}[!htb]
\centering
\begin{subfigure}[!htb]{0.5\textwidth}
    \centering
    \includegraphics[width=\textwidth]{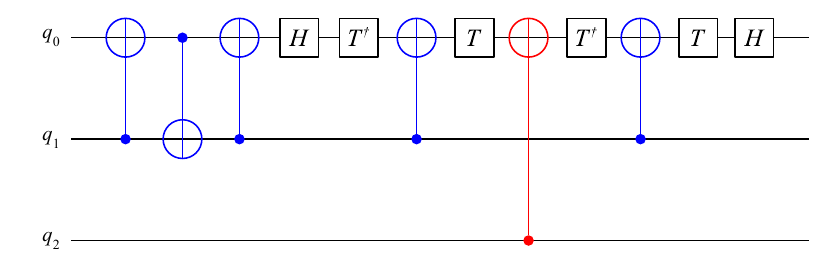}
    \caption{Decomposition with colors marking connections between qubits.}
    \label{fig:fig35a}
\end{subfigure}
\begin{subfigure}[!htb]{0.2\textwidth}
    \centering
    \includegraphics[width=\textwidth]{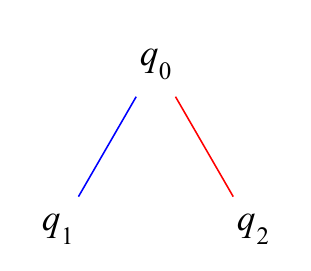}
    \caption{Connections between qubits.}
    \label{fig:fig35b}
\end{subfigure}
\caption{Decomposition and the connections between qubits in a single SWAT gate.}
\label{fig:fig35}
\end{figure}

As shown in Fig.~\ref{fig:fig35a}, the target qubit $q_0$ is always connected to both control qubits, $q_1$ (shown in \textcolor{blue}{blue}) and $q_2$ (shown in \textcolor{red}{red}). However, the controls $q_1$ and $q_2$ are never connected in a SWAT gate. Therefore, in a SWAT gate, to not have any additional global SWAP gates, the qubits should be connected as illustrated in Fig.~\ref{fig:fig35b}.

\begin{figure}[!htb]
\centering
\begin{subfigure}[!htb]{0.5\textwidth}
    \centering
    \includegraphics[width=\textwidth]{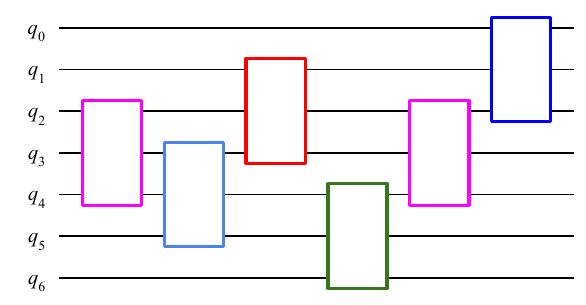}
\end{subfigure}
\begin{subfigure}[!htb]{0.3\textwidth}
    \centering
    \includegraphics[width=\textwidth]{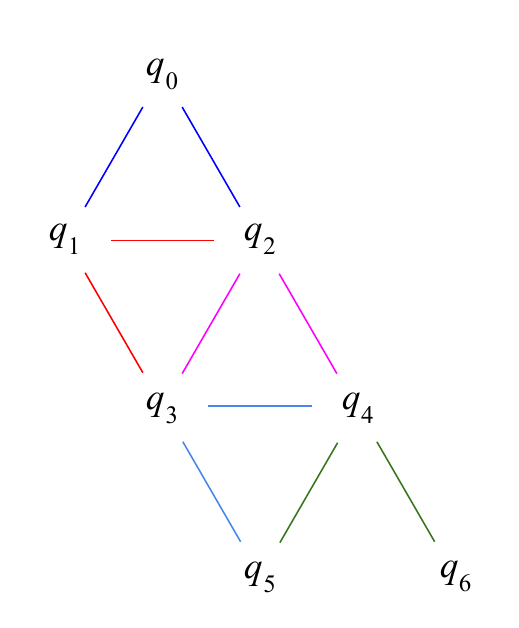}
\end{subfigure}
\caption{The triangular connectivity among qubits in a PDL-based circuit. Every colored block (a single SWAT gate) in the circuit on the left side is mapped to its corresponding connections in the triangular layout on the right side. Notice that the qubits on the left side are the logical qubits of a quantum circuit, while the qubits on the right side are the physical qubits of a quantum computer.}
\label{fig:fig36}
\end{figure}

\begin{figure}[!htb]
\centering
\includegraphics[width=0.3\textwidth]{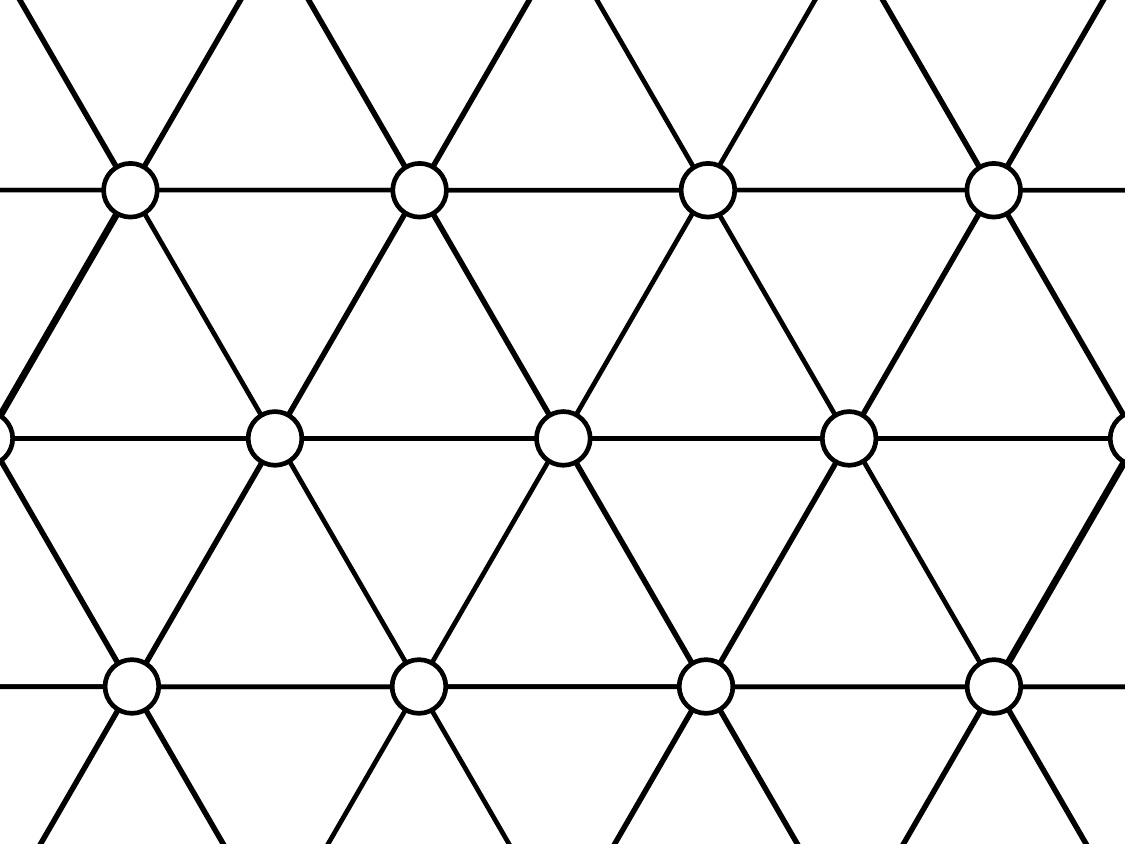}
\caption{Our proposed triangular layout for prospective quantum computers. Each physical qubit (represented by the white circles) has direct connections to 6 other neighboring qubits.}
\label{fig:fig23}
\end{figure}

We can use the structure in Fig.~\ref{fig:fig35b} to form the quantum layout by going through all the SWAT gates and finding the connections between qubits, as shown in Fig.~\ref{fig:fig36}.

From this approach, we introduce a new triangular quantum layout, as shown in Fig.~\ref{fig:fig23}, which is derived from the PDL-based circuit without additional global SWAP gates. This triangular layout is suitable for the Toffoli gate and lattice-based circuit synthesis techniques, leading to no additional global SWAP gates required.

\subsection{Square Grid and Heavy-Hex Mappings}
Although our triangular layout could potentially be a useful design, it is not currently fabricated. However, the connectivity of qubits in PDL-based circuits can also be mapped to other layouts, including the square grid \cite{arute, helmer, gong} and heavy-hex layout \cite{chamberland}. This section first covers how a linear mapping can be formed from the connectivity of qubits, and then shows how this can be mapped onto a square grid or heavy-hex layout.

\begin{figure}[!htb]
    \centering
    \includegraphics[width=0.3\linewidth]{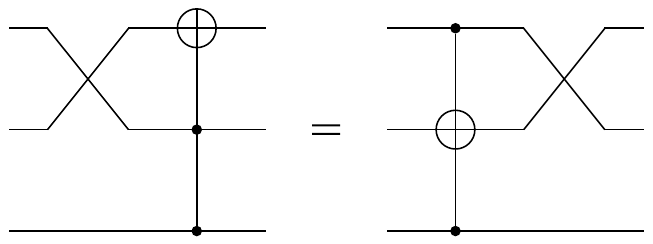}
    \caption{Two equivalent realizations of the SWAT gate.}
    \label{fig:swat-gate-v2}
\end{figure}

\begin{figure}[!htb]
    \centering
    \includegraphics[width=0.8\linewidth]{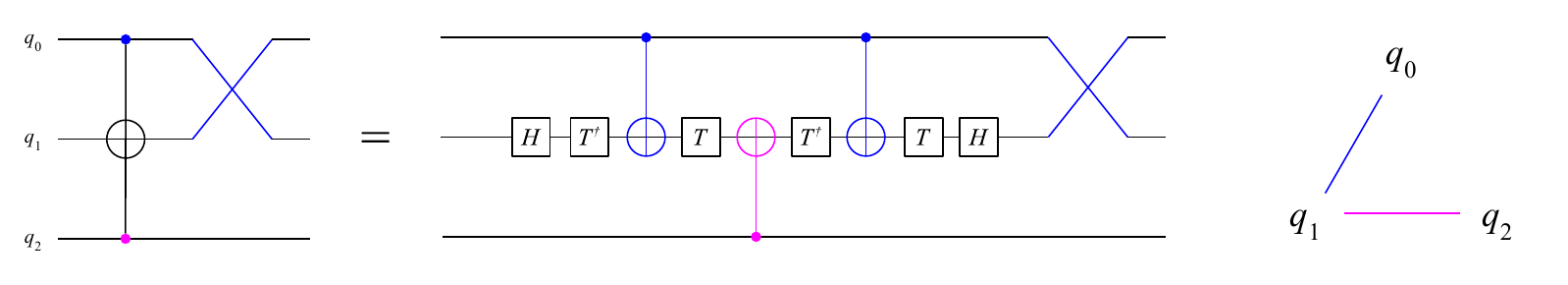}
    \caption{An equivalent decomposition qubit connectivity of the SWAT gate.}
    \label{fig:swat-decomp-v2}
\end{figure}

\begin{figure}[!htb]
    \centering
    \begin{subfigure}[!htb]{0.8\linewidth}
        \includegraphics[width=\linewidth]{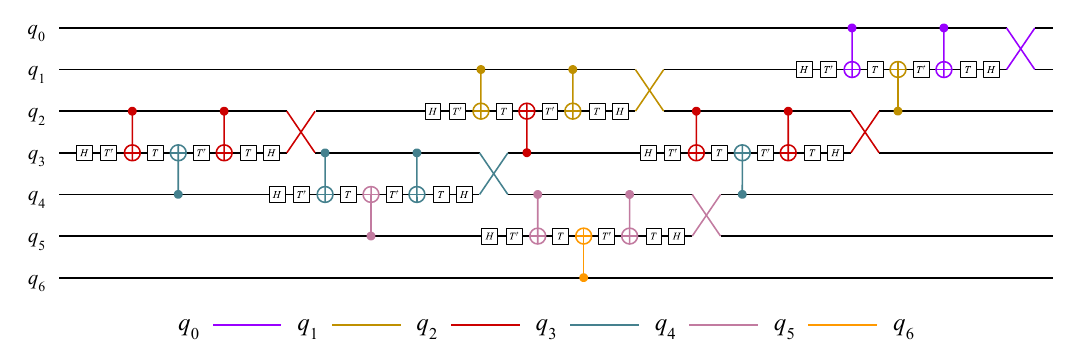}
        \caption{All connections between qubits in a PDL-based circuit with 3 levels based on the equivalent SWAT gate decomposition in Fig.~\ref{fig:swat-decomp-v2}. The color of a gate corresponds to its required connection among the qubits.}
        \label{fig:connectivity-v2}
    \end{subfigure}
    \centering
    \begin{subfigure}[!htb]{0.125\linewidth}
        \includegraphics[width=\linewidth]{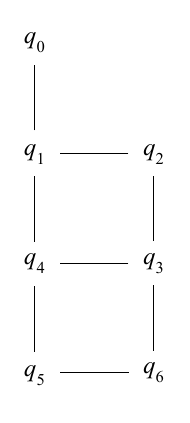}
        \caption{Square grid connectivity mapping.}
        \label{fig:square-grid-mapping}
    \end{subfigure}
    \begin{subfigure}[!htb]{0.4\linewidth}
        \includegraphics[width=\linewidth]{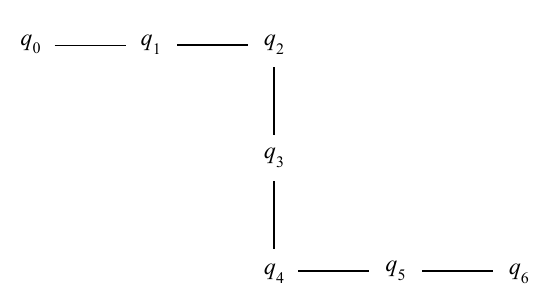}
        \caption{Heavy-hex connectivity mapping.}
        \label{fig:heavy-hex-mapping}
    \end{subfigure}
    \caption{The connectivity mappings among the qubits in a 3-level PDL-based circuit.}
\end{figure}

The SWAT gate can be redesigned as shown in Fig.~\ref{fig:swat-gate-v2}, where the Toffoli gate is applied before the SWAP Gate. Rewriting the gate in this way leads to a different connectivity between qubits, as illustrated in Fig.~\ref{fig:swat-decomp-v2}.

Thus, the qubit layout shown in Fig.~\ref{fig:connectivity-v2} for a 3-level PDL-based circuit can be constructed by finding the best qubit connectivity for each SWAT gate based on Fig.~\ref{fig:swat-decomp-v2}, leading to a linear connectivity mapping. This results in a mapping that reduces the number of connections among qubits with no global SWAP gates. This is the case because, apart from the local SWAP gates contained inside the SWAT gates, all the other gates, specifically two-qubit gates, act on connected qubits, so no global SWAP gates are necessary.

This qubit layout can be rearranged to fit both the square grid and heavy-hex layouts, as shown in Fig.~\ref{fig:square-grid-mapping} and Fig.~\ref{fig:heavy-hex-mapping}, respectively, without additional global SWAP gates required.

\subsection{Cost of PDL-Based Circuit For Symmetric Functions}
In this section, we calculate the cost of a PDL-based circuit for an $n$-variable symmetric function in the worst case, where the whole $n$-level lattice must be built.

To compare the costs of different circuits, two cost systems will be used: Maslov cost~\cite{maslov} and Transpilation Quantum Cost (TQC)~\cite{albayaty-gala, albayaty-cala}.

The Maslov cost is a hardware-independent approach that sums up the individual cost of all the different Toffoli gates depending on the number of controls.

In the worst case, an $n$-level lattice contains $\frac{n(n+1)}{2}$ SWAT gates, so there are $\frac{n(n+1)}{2}$ SWAP gates and $\frac{n(n+1)}{2}$ 3-bit Toffoli gates. A single SWAP gate is composed of 3 CNOT gates, which amounts to a cost of $\frac{3n(n+1)}{2}$. Each CNOT gate has a Maslov cost of 1 (total cost of $\frac{3n(n+1)}{2}$ from CNOT gates), and each 3-bit Toffoli gate has a Maslov cost of 5 (total cost of $\frac{5n(n+1)}{2}$ from Toffoli gates), thus the total Maslov cost of an $n$-level lattice, in the worst case, is $4n(n+1)$. All of this is summarized in Table~\ref{tab:pdl-maslov-cost}.

\begin{table}[!htb]
    \centering
    \renewcommand{\arraystretch}{1.5}
    \resizebox{0.6\linewidth}{!}{
    \begin{tabular}{c||cc|c}
         & CNOT Gate & 3-bit Toffoli Gate & \\
         \cline{1-3}
         Amount & $\frac{3n(n+1)}{2}$ & $\frac{n(n+1)}{2}$ \\
         Individual Cost & 1 & 5 & Total Cost \\
         \hline
         Cost & $\frac{3n(n+1)}{2}$ & $\frac{5n(n+1)}{2}$ & $4n(n+1)$
    \end{tabular}}
    \caption{Gates and Maslov cost of an $n$-level PDL-based circuit.}
    \label{tab:pdl-maslov-cost}
\end{table}

\begin{figure}[!htb]
    \centering
    \includegraphics[width=0.7\linewidth]{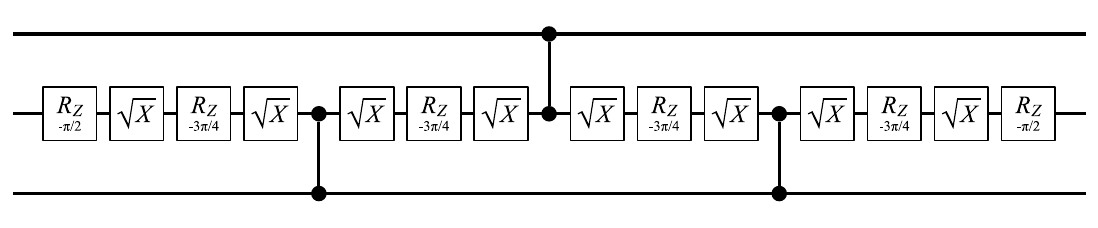}
    \caption{The SWAT gate decomposed into native gates using the IBM Qiskit Transpiler.}
    \label{fig:transpiled-swat}
\end{figure}

The TQC is a quantum cost that calculates the cost of a final transpiled circuit. This cost was originally designed for IBM QPUs, but it can be used for any hardware. The TQC of a circuit is determined by~\eqref{eqn:tqc}, where $N_1$ is the number of native single-qubit gates, $N_2$ is the number of native double-qubit gates, $XC$ is the number of SWAP gates, and $D$ is the depth. The TQC in this paper is calculated based on IBM native gates.
\begin{equation}
    \label{eqn:tqc}
    \text{TQC} = N_1 + N_2 + XC + D
\end{equation}
The SWAT gate from Fig.~\ref{fig:swat-decomp-v2} can be further decomposed into the IBM native gates as shown in Fig.~\ref{fig:transpiled-swat}, which is done here by using the CALA-$n$ decomposition and IBM Qiskit Transpiler. 

\section{Results and Discussion}
\begin{figure}[!htb]
    \centering
    \begin{subfigure}[!htb]{0.41\linewidth}
        \centering
        \includegraphics[width=\linewidth]{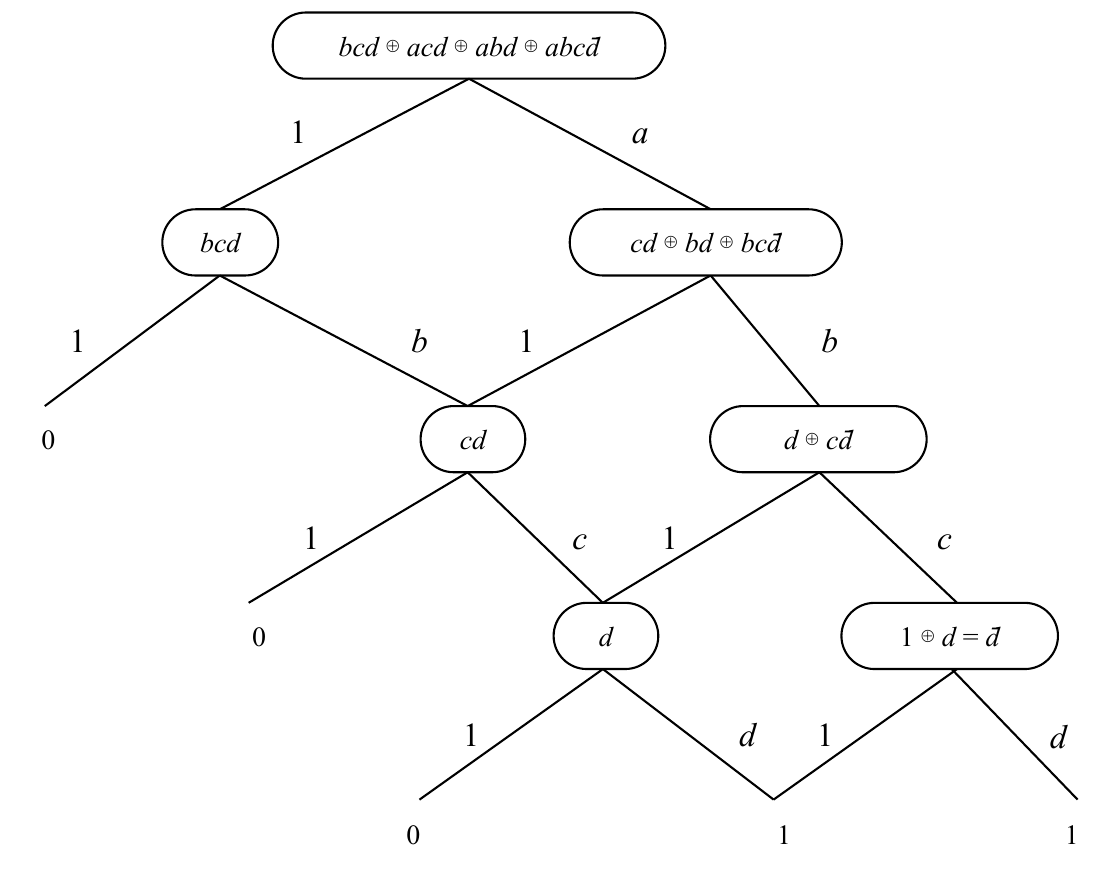}
        \caption{PDL.}
        \label{fig:example1-lattice}
    \end{subfigure}
    \begin{subfigure}[!htb]{0.3\linewidth}
        \centering
        \includegraphics[width=\linewidth]{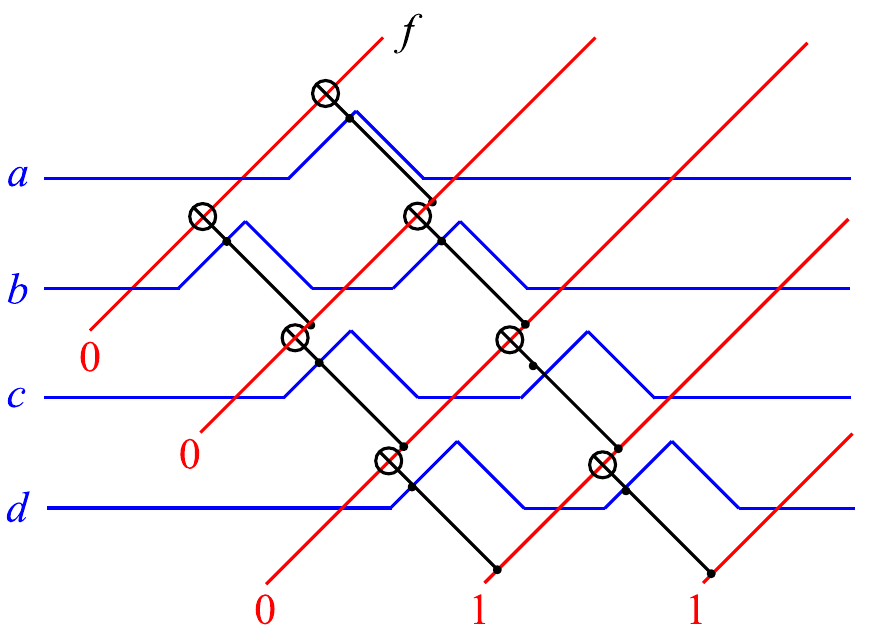}
        \caption{Circuit-lattice.}
        \label{fig:example1-circuit-lattice}
    \end{subfigure}
    \begin{subfigure}[!htb]{0.3\linewidth}
        \centering
        \includegraphics[width=\linewidth]{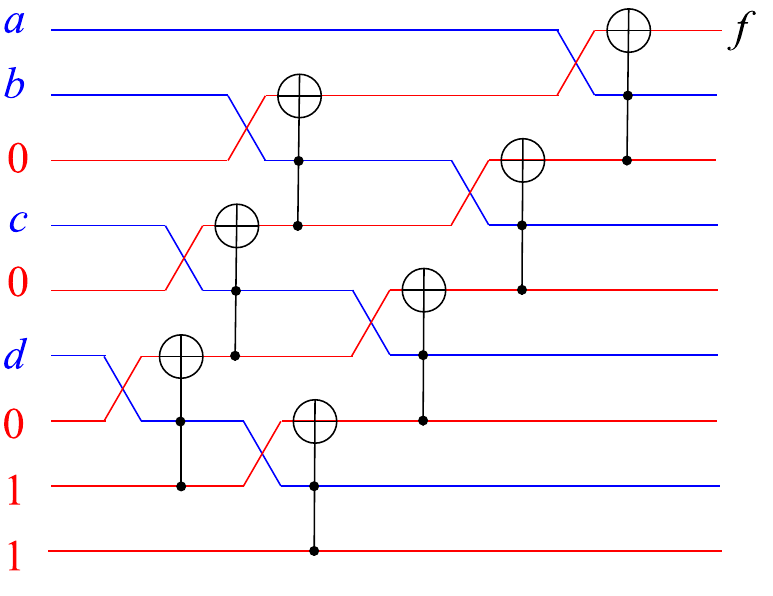}
        \caption{PDL-based circuit.}
        \label{fig:example1-pdl-circuit}
    \end{subfigure}
    \begin{subfigure}[!htb]{0.2\linewidth}
        \centering
        \includegraphics[width=\linewidth]{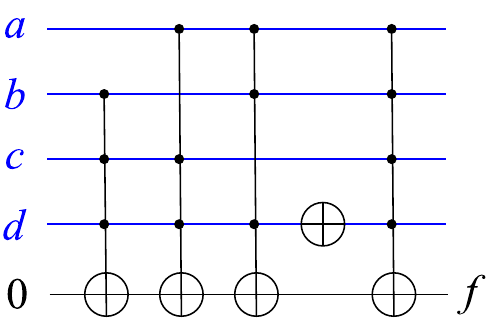}
        \caption{Circuit realized directly from the ESOP expression.}
        \label{fig:example1-esop-circuit}
    \end{subfigure}
    \caption{Example 1 of PDL and circuits with the function $f = bcd \oplus acd \oplus abd \oplus abc\bar{d}$.}
\end{figure}

ESOP expressions are commonly used for quantum circuit synthesis~\cite{schmitt, wille}. However, synthesizing a circuit directly from an ESOP expression is cost-expensive, requiring a larger number of gates, and PDL-based circuits are less costly in many cases. This section presents multiple comparison examples of these cases to discuss the benefits of PDL-based circuits.

\noindent 
\\\textbf{Comparison Example 1:} $f = bcd \oplus acd \oplus abd \oplus abc\bar{d}$. The PDL for this function is drawn in Fig.~\ref{fig:example1-lattice}. To create a circuit for this function, the circuit-lattice diagram is first created, as shown in Fig~\ref{fig:example1-circuit-lattice}.

This circuit-lattice is used to construct the circuit, as illustrated in Fig.~\ref{fig:example1-pdl-circuit}. This circuit has a Maslov cost~\cite{maslov} of 56 and a TQC of 277 or 236 (with CALA-$n$, the decomposition from Fig. \ref{fig:swat-decomp-v2}) after using the IBM Qiskit Transpiler \cite{qiskit}. If the circuit was realized directly from the ESOP expression, as shown in Fig.~\ref{fig:example1-esop-circuit}, then it would have a Maslov cost of 69 and a TQC of 1048 instead, without factoring in the additional global SWAP gates required from the layout. Using our PDL-based circuit synthesis method, we gain at least 13 Maslov cost and, together with CALA-$n$, 812 TQC.

\noindent 
\\\textbf{Comparison Example 2:} $f = e \oplus d \oplus c \oplus cde \oplus b \oplus bde \oplus bce \oplus bcd\bar{e} \oplus a \oplus ade \oplus ace \oplus acd\bar{e} \oplus abe \oplus \bar{e} \oplus abc\bar{d}\bar{e}$.

\begin{figure}[!htb]
    \centering
    \begin{subfigure}[!htb]{0.9\linewidth}
        \centering
        \includegraphics[width=\linewidth]{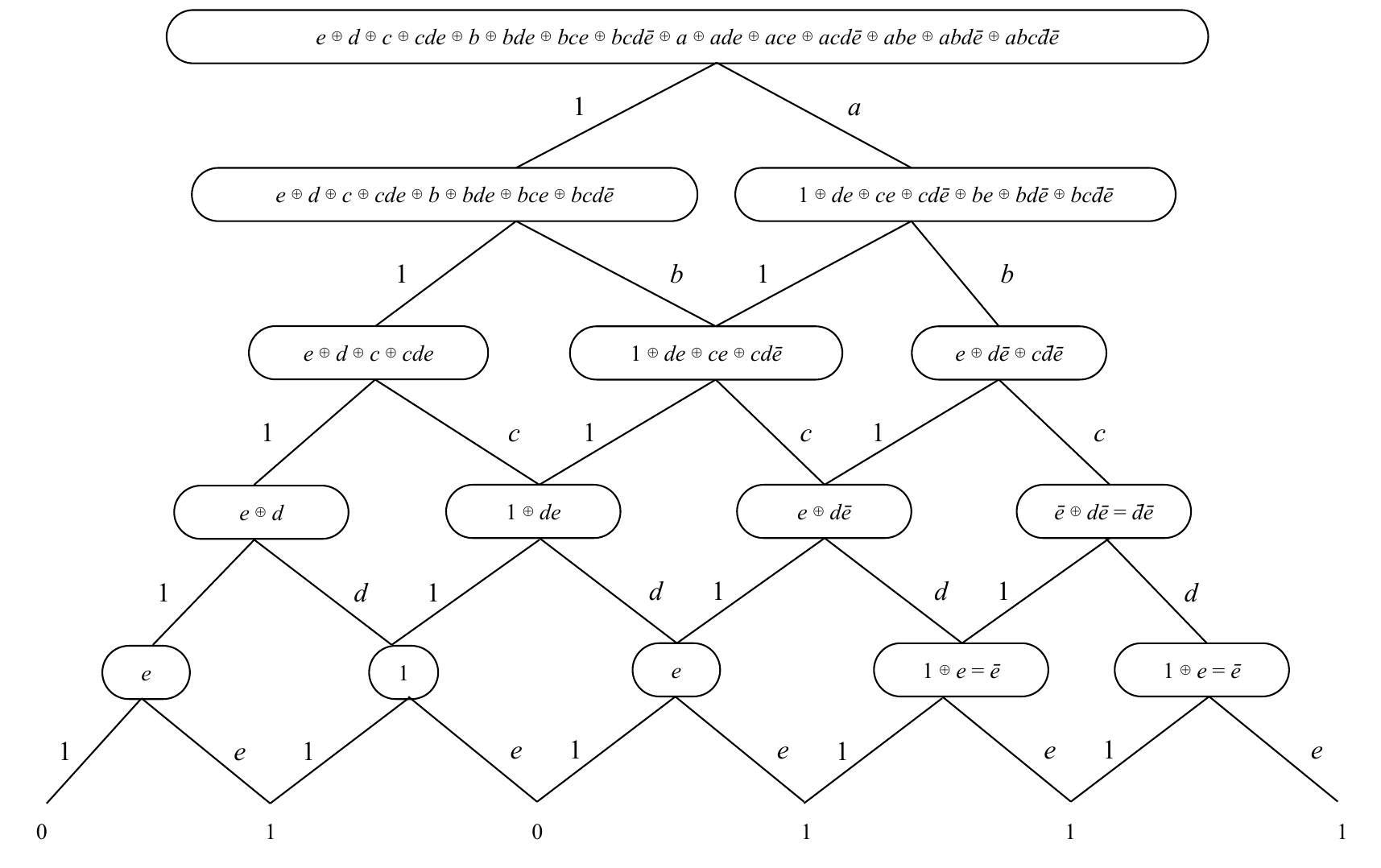}
        \caption{PDL.}
        \label{fig:example2-lattice}
    \end{subfigure}
    \begin{subfigure}[!htb]{0.35\linewidth}
        \centering
        \includegraphics[width=\linewidth]{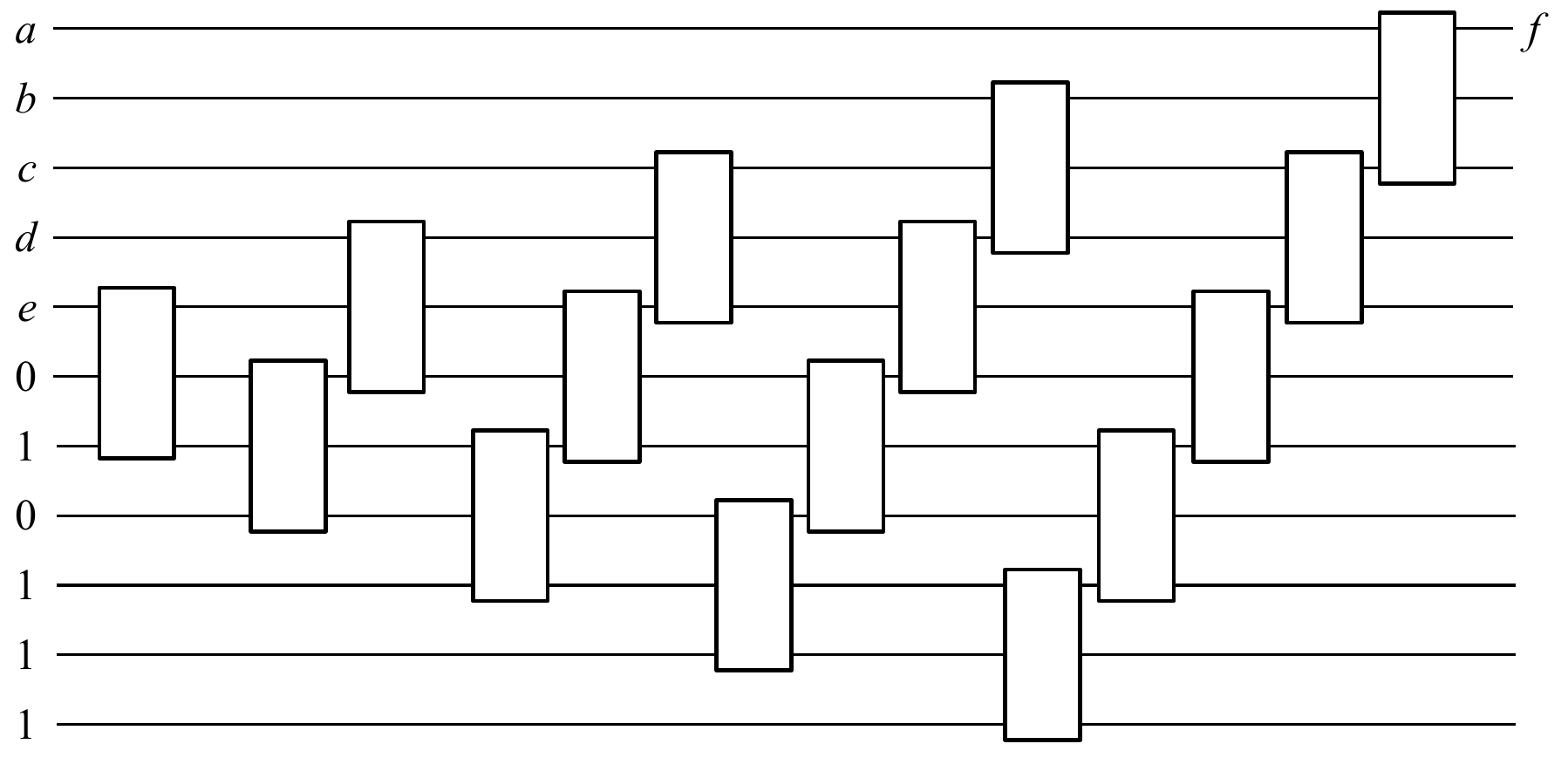}
        \caption{PDL-based circuit, where the rectangles represent SWAT gates.}
        \label{fig:example2-pdl-circuit}
    \end{subfigure}
    \begin{subfigure}[!htb]{0.5\linewidth}
        \centering
        \includegraphics[width=\linewidth]{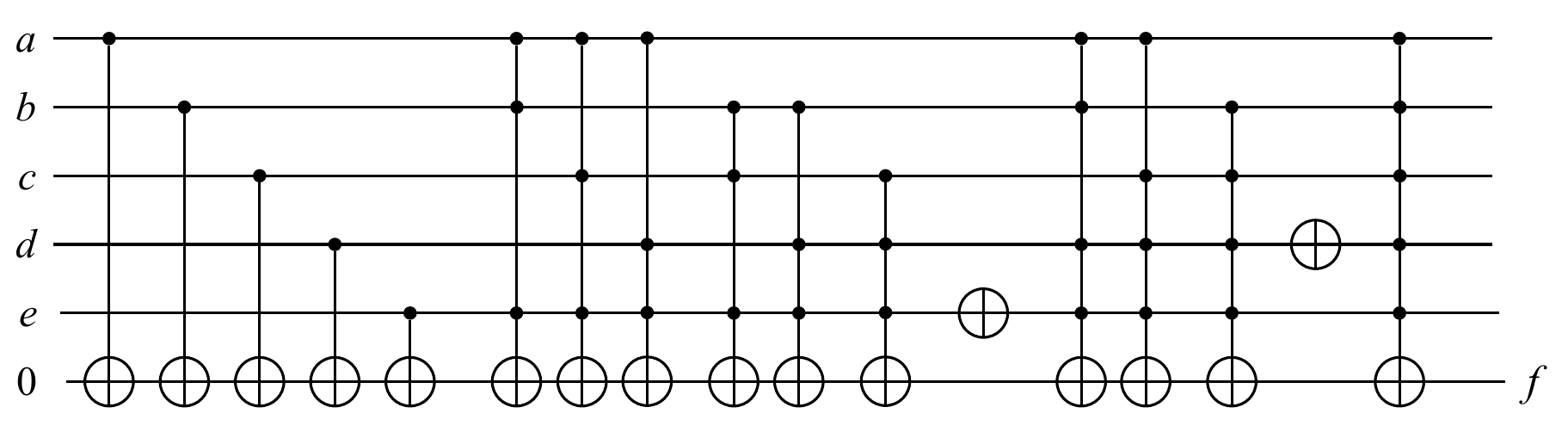}
        \caption{Circuit realized directly from the ESOP expression.}
        \label{fig:example2-esop-circuit}
    \end{subfigure}
    \caption{Example 2 of PDL and circuits with the function $f = e \oplus d \oplus c \oplus cde \oplus b \oplus bde \oplus bce \oplus bcd\bar{e} \oplus a \oplus ade \oplus ace \oplus acd\bar{e} \oplus abe \oplus \bar{e} \oplus abc\bar{d}\bar{e}$.}
\end{figure}

The PDL for this function is shown in Fig.~\ref{fig:example2-lattice}. The PDL-based circuit is illustrated in Fig.~\ref{fig:example2-pdl-circuit}, and it has a Maslov cost of 120 and, with CALA-$n$, a TQC of 490, while the ESOP-based circuit is shown in Fig.~\ref{fig:example2-esop-circuit}, and it has a Maslov cost of at least 215, without including the additional global SWAP gates necessary for the layout, and a TQC of 3709. Thus, by using the PDL-based circuit synthesis, we obtain at least 95 Maslov cost and 3219 TQC (when used with CALA-$n$).

\noindent 
\\\textbf{Comparison Example 3:} $f = e \oplus d \oplus c \oplus cde \oplus b \oplus bde \oplus bce \oplus bcd \oplus a \oplus ade \oplus ace \oplus acd \oplus abe \oplus abd \oplus abc \oplus abcde$.

\begin{figure}[!htb]
    \centering
    \begin{subfigure}[!htb]{0.6\linewidth}
        \centering
        \includegraphics[width=\linewidth]{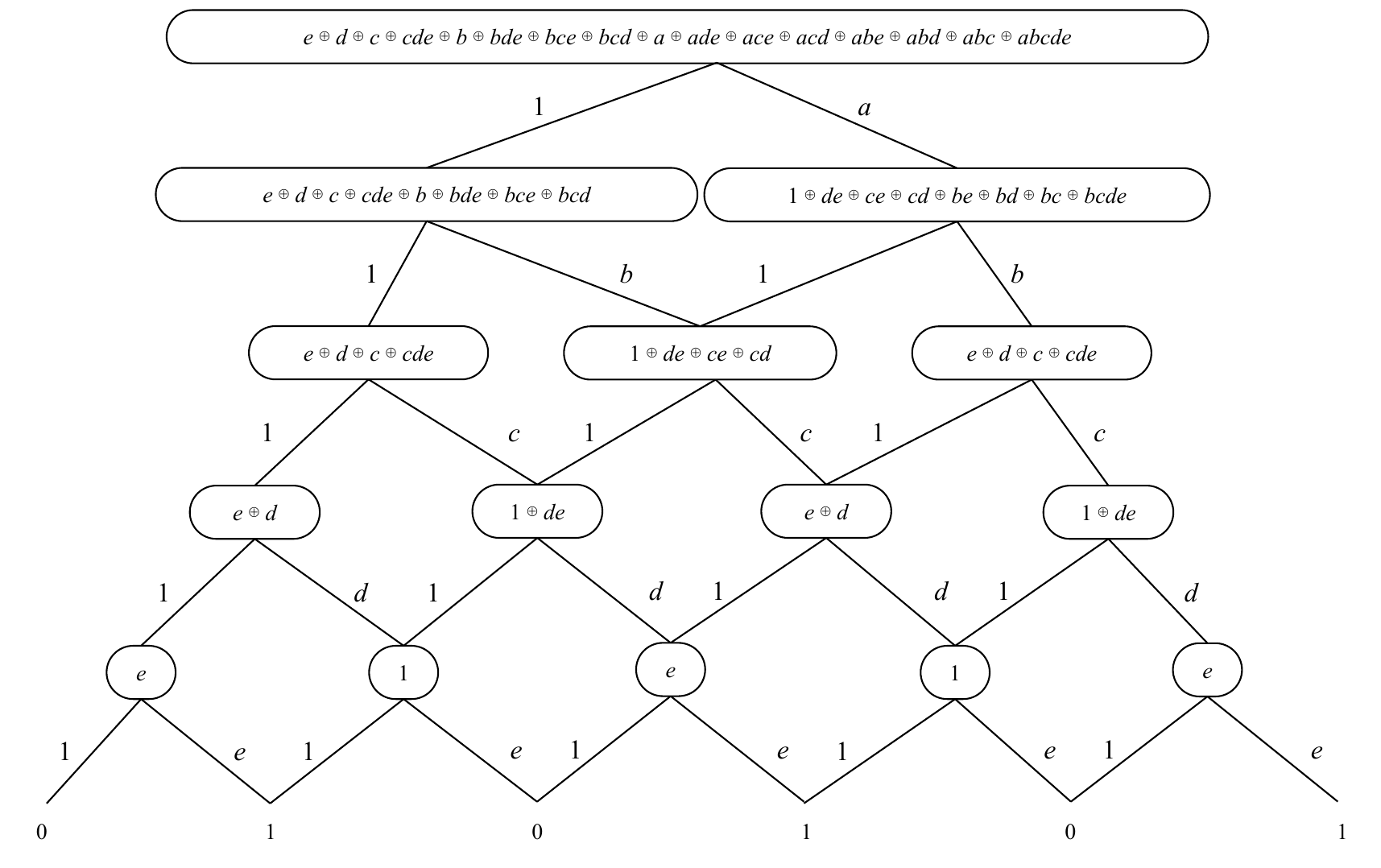}
        \caption{PDL for the $f$.}
        \label{fig:example3-lattice}
    \end{subfigure}
    \begin{subfigure}[!htb]{0.35\linewidth}
        \centering
        \includegraphics[width=\linewidth]{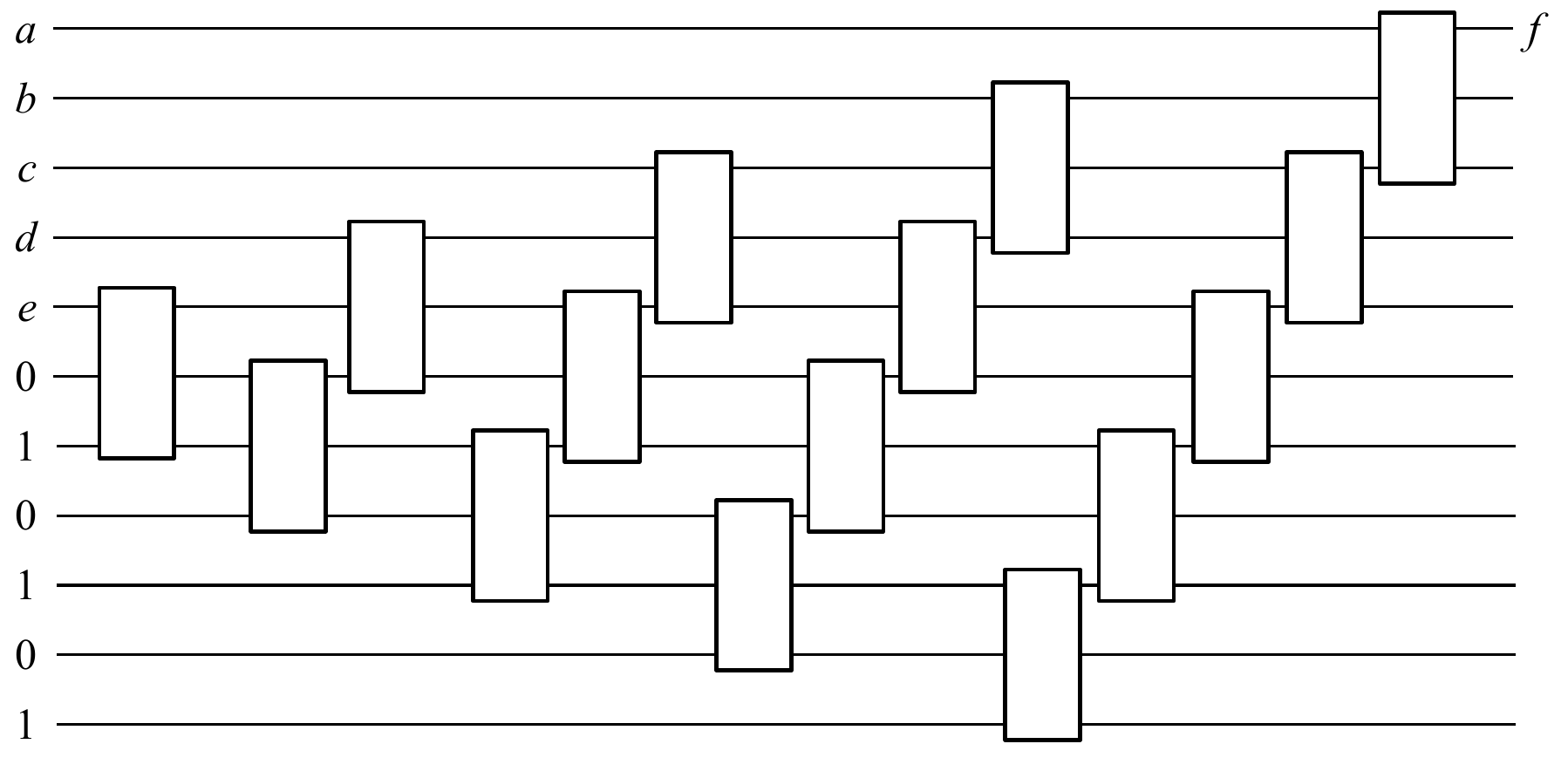}
        \caption{PDL-based circuit for $f$, where the rectangles represent SWAT gates.}
        \label{fig:example3-pdl-circuit}
    \end{subfigure}
    \begin{subfigure}[!htb]{0.5\linewidth}
        \centering
        \includegraphics[width=\linewidth]{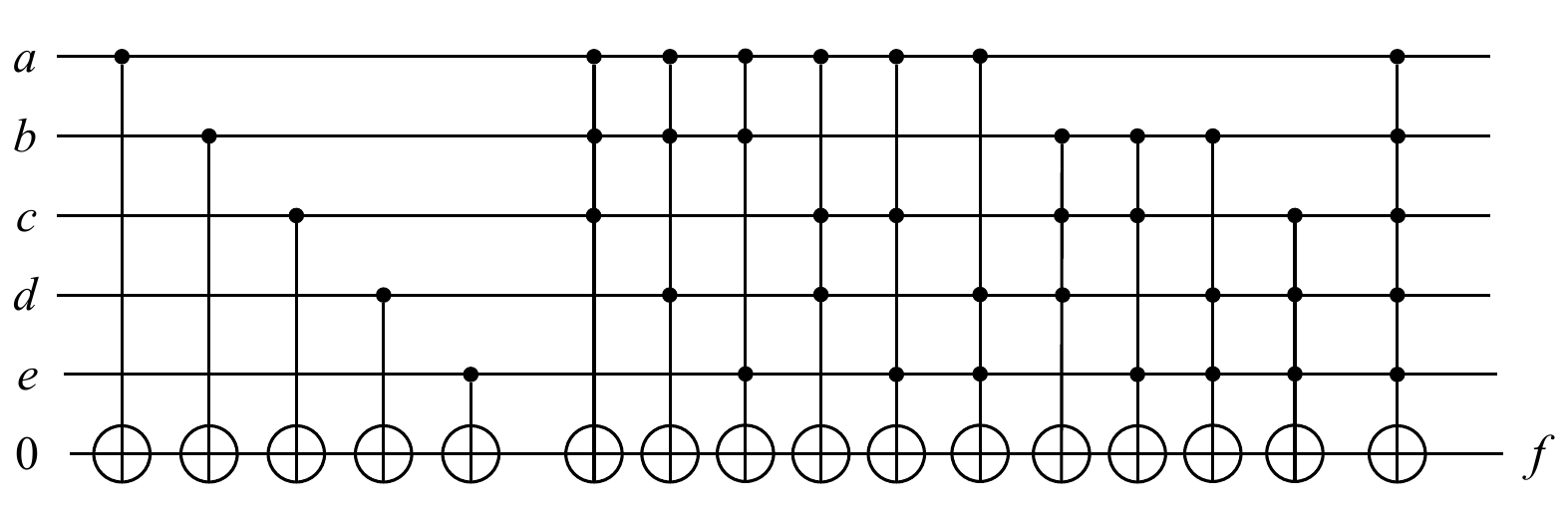}
        \caption{Circuit for $f$ realized directly from the ESOP expression.}
        \label{fig:example3-esop-circuit}
    \end{subfigure}
    \caption{Example 3 of PDL and circuits with the function $f = e \oplus d \oplus c \oplus cde \oplus b \oplus bde \oplus bce \oplus bcd \oplus a \oplus ade \oplus ace \oplus acd \oplus abe \oplus abd \oplus abc \oplus abcde$.}
\end{figure}

A possible PDL, shown in Fig.~\ref{fig:example3-lattice}, which leads to the PDL-based circuit in Fig.~\ref{fig:example3-pdl-circuit} with a Maslov cost of 120 and TQC of 489 (when used with CALA-$n$). The ESOP-based circuit, shown in Fig.~\ref{fig:example3-esop-circuit}, has a Maslov cost of at least 183 (excluding additional global SWAP gates) and TQC of 3331. Therefore, utilizing the PDL-based circuit synthesis method leads to a gain of at least 63 Maslov cost and 2842 (with CALA-$n$).

\noindent 
\\\textbf{Comparison Example 4:} $f = u \oplus av \oplus bv \oplus cv \oplus dv \oplus ev \oplus abw \oplus acw \oplus adw \oplus aew \oplus bcw \oplus bdw \oplus bew \oplus cdw \oplus dew \oplus abcx \oplus abdx \oplus abex \oplus acdx \oplus acex \oplus adex \oplus bcdx \oplus bcex \oplus bdex \oplus cdex \oplus abcdy \oplus abcey \oplus abdey \oplus acdey \oplus bcdey \oplus abcdez$.

\begin{figure}[!htb]
    \centering
    \begin{subfigure}[!htb]{0.43\linewidth}
        \centering
        \includegraphics[width=\linewidth]{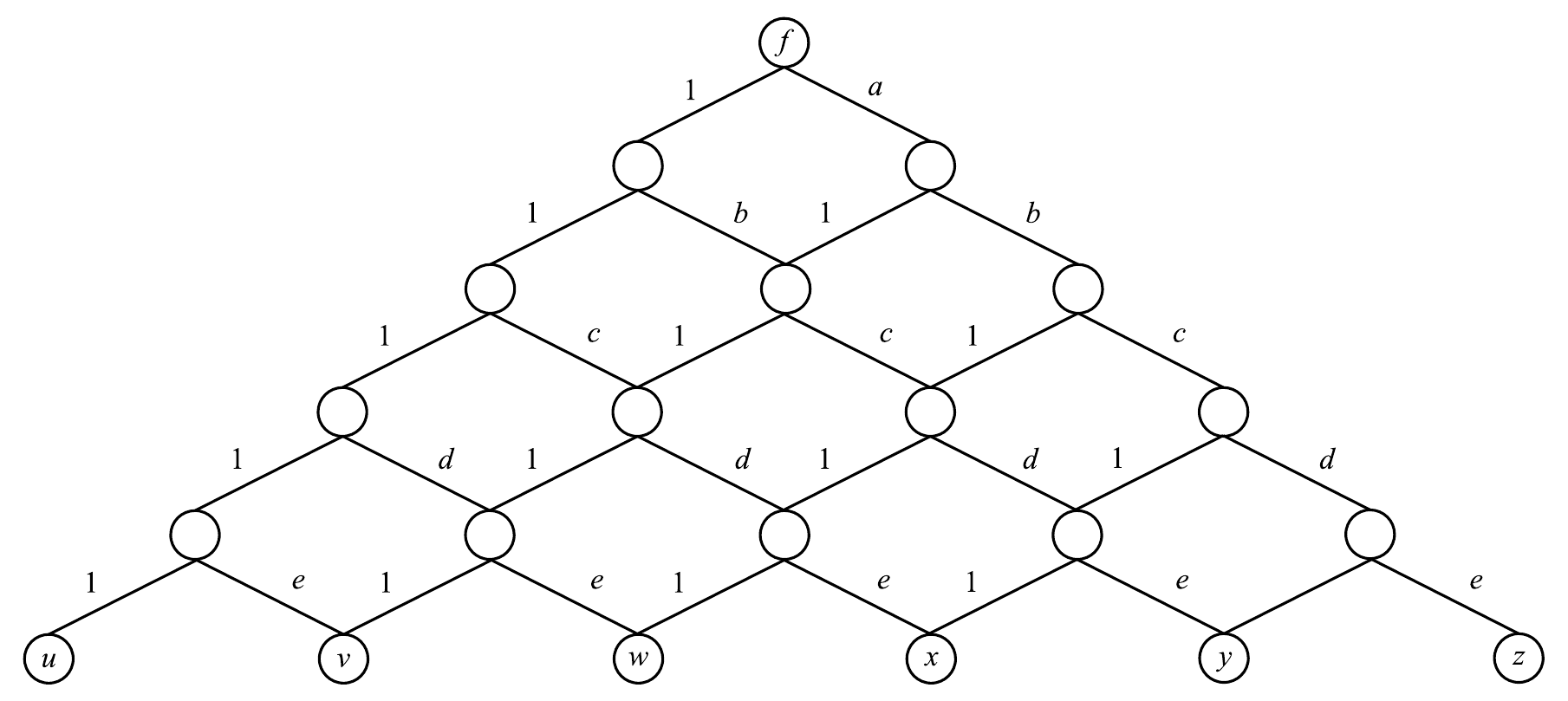}
        \caption{PDL.}
        \label{fig:example4-lattice}
    \end{subfigure}
    \begin{subfigure}[!htb]{0.4\linewidth}
        \centering
        \includegraphics[width=0.9\linewidth]{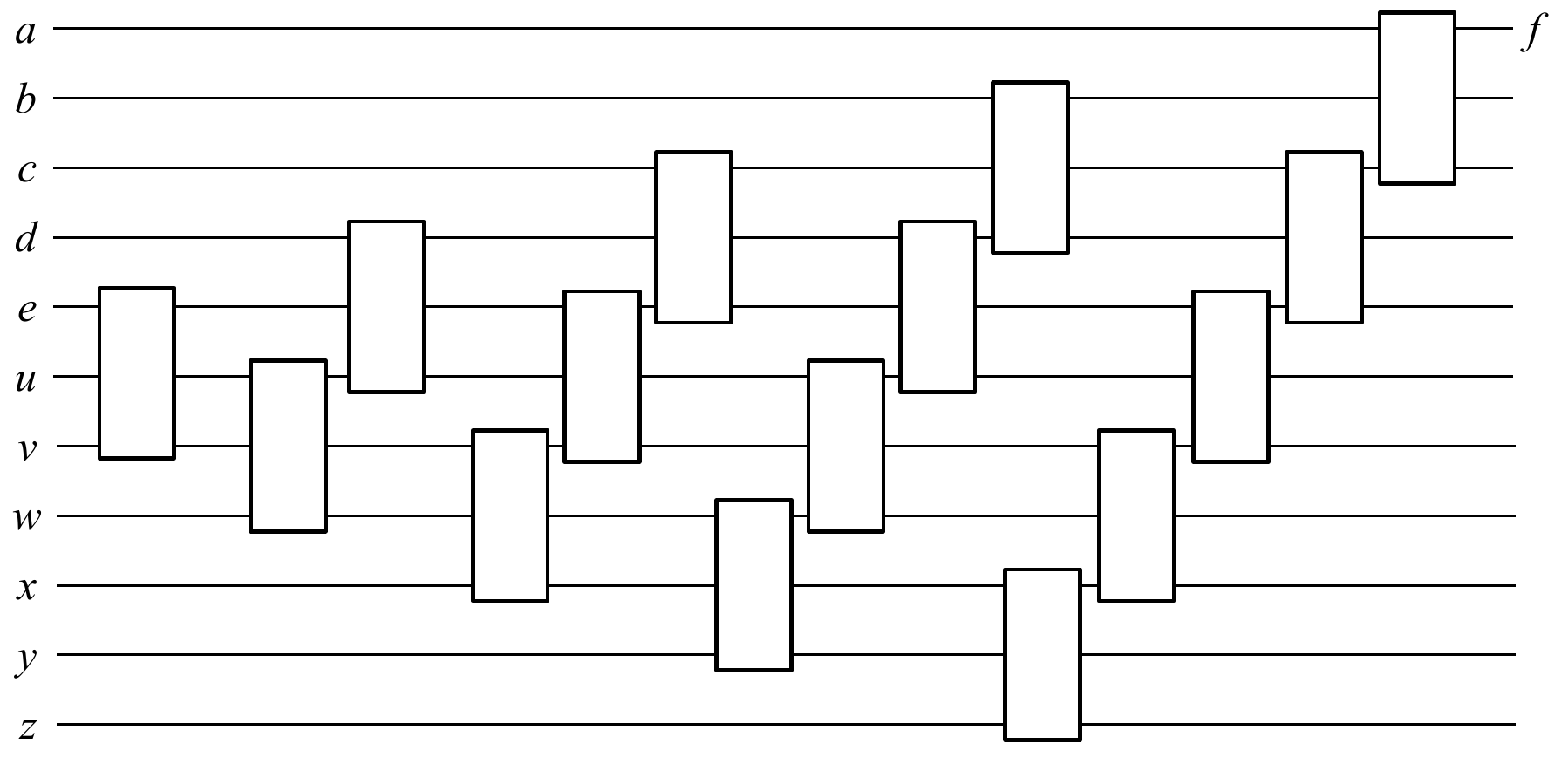}
        \caption{PDL-based circuit, where the rectangles represent SWAT gates.}
        \label{fig:example4-pdl-circuit}
    \end{subfigure}
    \begin{subfigure}[!htb]{0.6\linewidth}
        \centering
        \includegraphics[width=\linewidth]{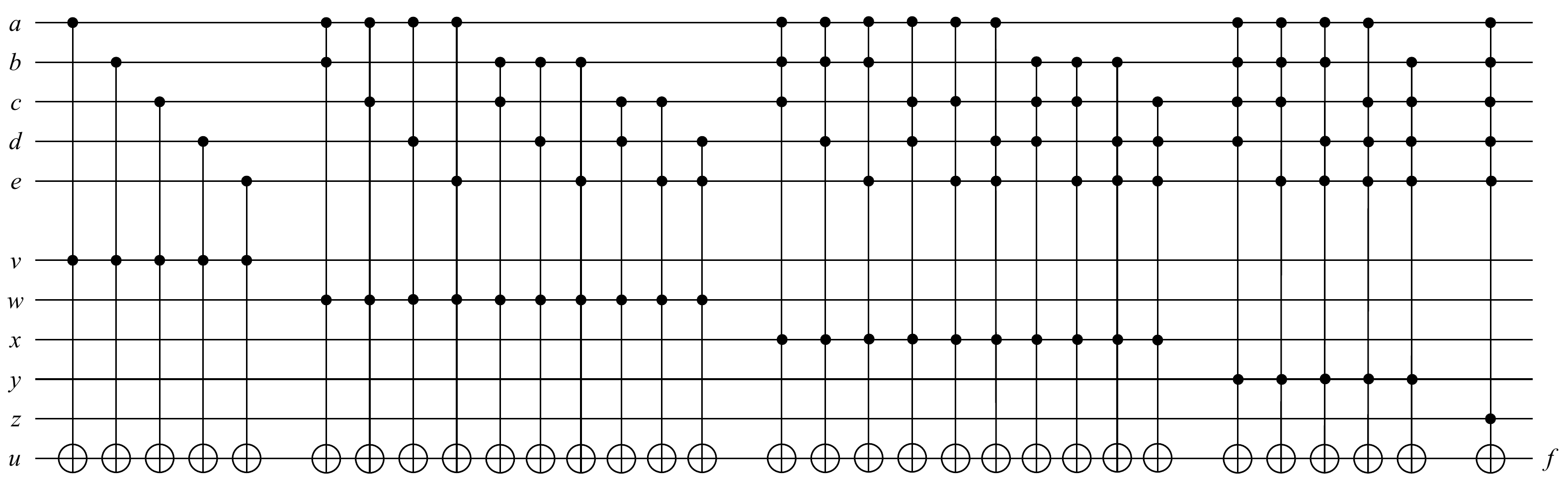}
        \caption{Circuit realized directly from the ESOP expression.}
        \label{fig:example4-esop-circuit}
    \end{subfigure}
    \caption{Example 4 of PDL and circuits with the function $f = u \oplus av \oplus bv \oplus cv \oplus dv \oplus ev \oplus abw \oplus acw \oplus adw \oplus aew \oplus bcw \oplus bdw \oplus bew \oplus cdw \oplus dew \oplus abcx \oplus abdx \oplus abex \oplus acdx \oplus acex \oplus adex \oplus bcdx \oplus bcex \oplus bdex \oplus cdex \oplus abcdy \oplus abcey \oplus abdey \oplus acdey \oplus bcdey \oplus abcdez$.}
\end{figure}

The PDL for this function is shown in Fig.~\ref{fig:example4-lattice}. The PDL-based circuit, which has a Maslov cost of 120 (excluding additional global SWAP gates) and TQC of 493 (when used with CALA-$n$), is shown in Fig.~\ref{fig:example4-pdl-circuit}, and the ESOP-based circuit, as shown in Fig.~\ref{fig:example4-esop-circuit}, has a Maslov cost of at least 750 (excluding additional global SWAP gates) and TQC of 8876. Therefore, utilizing the PDL-based circuit synthesis method leads to a gain of at least 630 Maslov cost and 8383 (with CALA-$n$).

Note that this function is not fully symmetric: $a$, $b$, $c$, $d$, and $e$ can be interchanged with each other and yield the same function, but this is not the case with any other pairs. Thus, the function is $\frac{2}{11} \approx 18.182 \%$ symmetric.

\begin{figure}[!htb]
    \centering
    \begin{subfigure}[!htb]{0.65\linewidth}
        \centering
        \includegraphics[width=\linewidth]{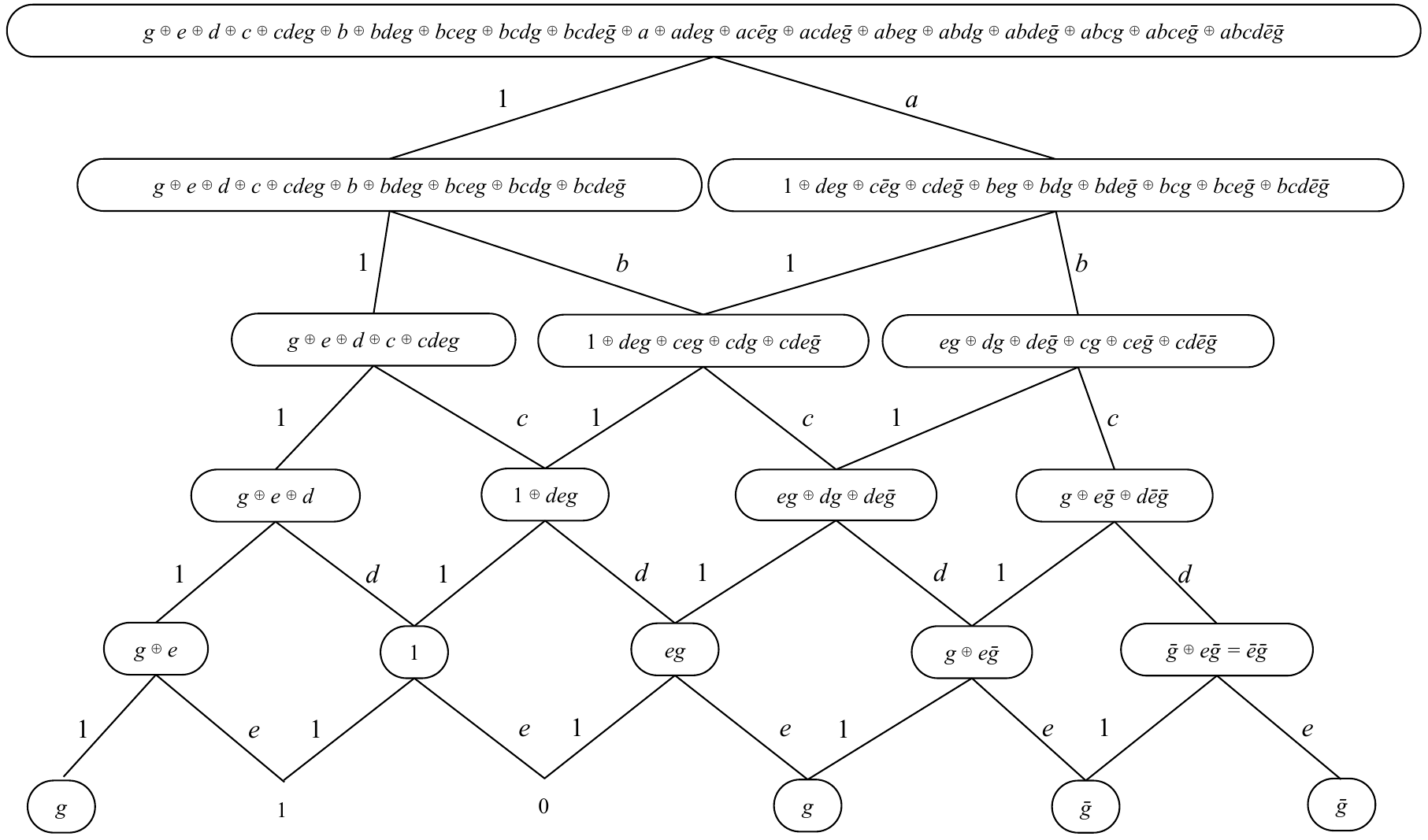}
        \caption{PDL for $f$.}
        \label{fig:example5-lattice}
    \end{subfigure}
    \begin{subfigure}[!htb]{0.42\linewidth}
        \centering
        \includegraphics[width=\linewidth]{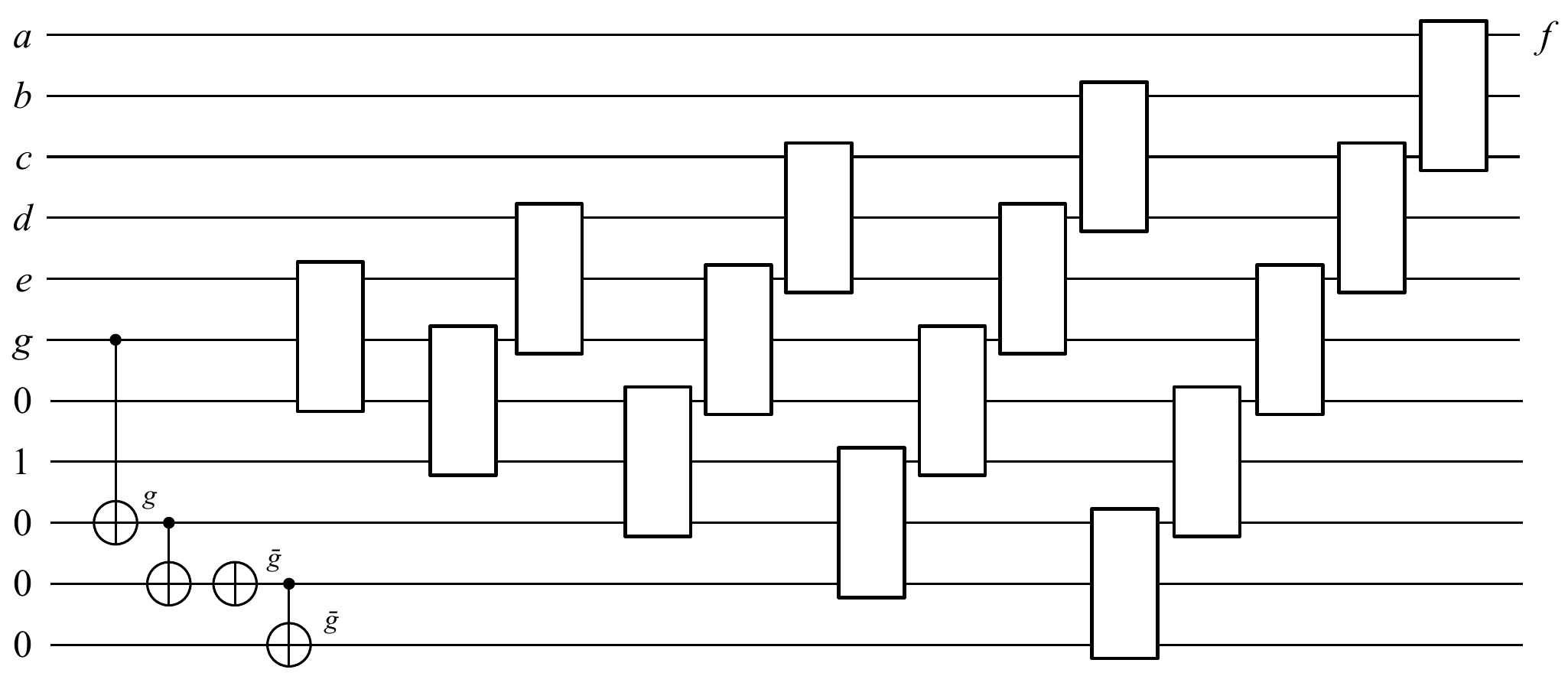}
        \caption{PDL-based circuit for $f$, where the rectangles represent SWAT gates.}
        \label{fig:example5-pdl-circuit}
    \end{subfigure}
    \begin{subfigure}[!htb]{0.5\linewidth}
        \centering
        \includegraphics[width=\linewidth]{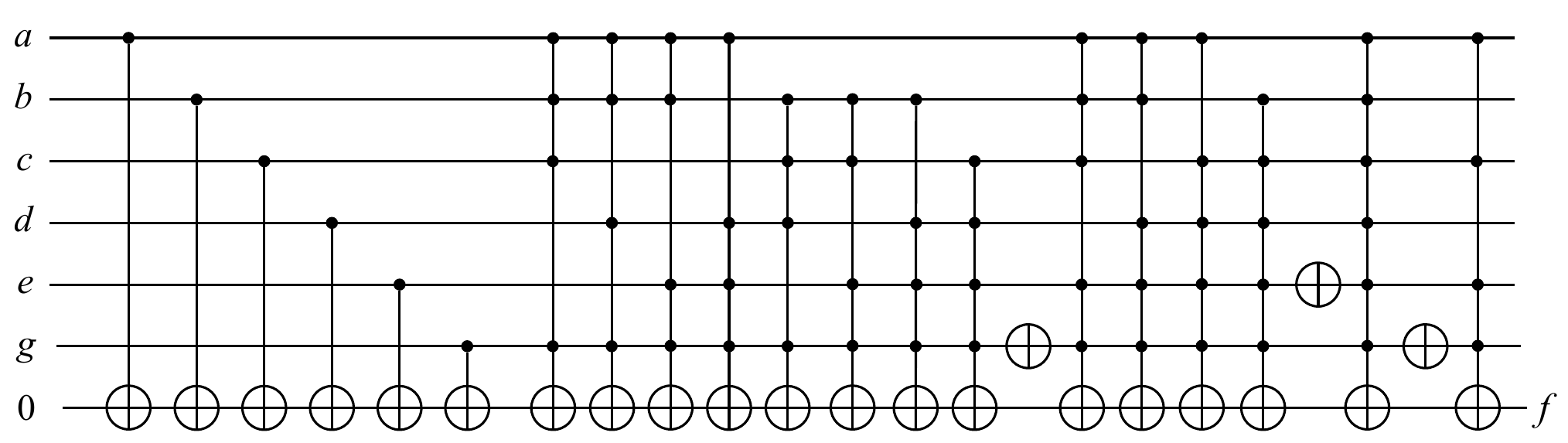}
        \caption{Circuit for $f$ realized directly from the ESOP expression.}
        \label{fig:example5-esop-circuit}
    \end{subfigure}
    \caption{Example 5 of PDL and circuits with the function $f = g \oplus e \oplus d \oplus c \oplus cdeg \oplus b \oplus bdeg \oplus bceg \oplus bcdg \oplus bcde\bar{g} \oplus a \oplus adeg \oplus ac\bar{e}g \oplus acde\bar{g} \oplus abeg \oplus abdg \oplus abde\bar{g} \oplus abcg \oplus abce\bar{g} \oplus abcd\bar{e}\bar{g}$.}
\end{figure}

\noindent 
\\\textbf{Comparison Example 5:} $f = g \oplus e \oplus d \oplus c \oplus cdeg \oplus b \oplus bdeg \oplus bceg \oplus bcdg \oplus bcde\bar{g} \oplus a \oplus adeg \oplus ac\bar{e}g \oplus acde\bar{g} \oplus abeg \oplus abdg \oplus abde\bar{g} \oplus abcg \oplus abce\bar{g} \oplus abcd\bar{e}\bar{g}$

The PDL for $f$ that is shown in Fig.~\ref{fig:example5-lattice} corresponds to the PDL-based circuit in Fig.~\ref{fig:example5-pdl-circuit}, which has a Maslov cost of 124 and TQC of 520 (when used with CALA-$n$). The ESOP-based circuit for $f$, shown in Fig.~\ref{fig:example5-esop-circuit}, has a Maslov cost of at least 547 (excluding additional global SWAP gates) and TQC of 9776. Thus, the PDL-based circuit synthesis method leads to a gain of at least 423 Maslov cost and 9256 (with CALA-$n$).

\begin{figure}[!htb]
    \centering
    \begin{subfigure}[!htb]{0.4\linewidth}
        \centering
        \includegraphics[width=\linewidth]{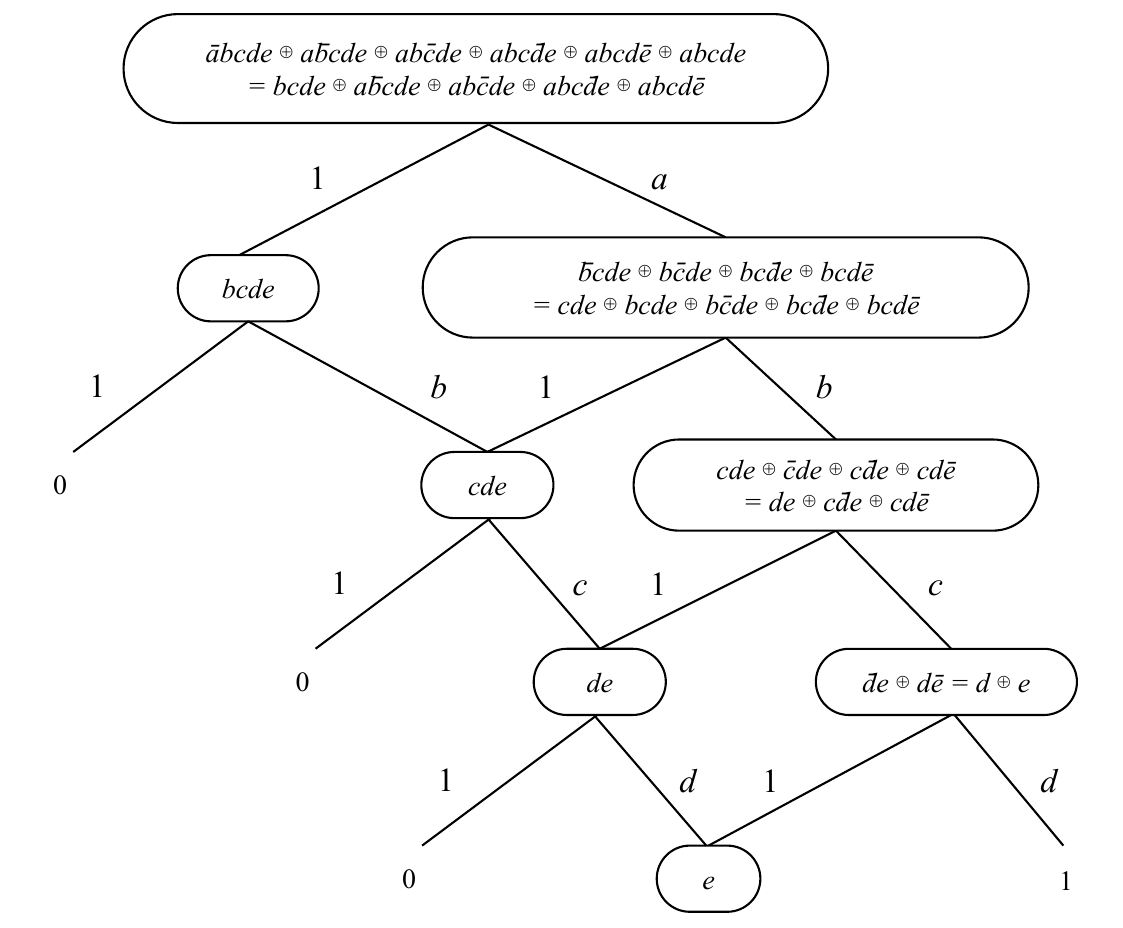}
        \caption{PDL for $f$.}
        \label{fig:example6-lattice}
    \end{subfigure}
    \begin{subfigure}[!htb]{0.3\linewidth}
        \centering
        \includegraphics[width=\linewidth]{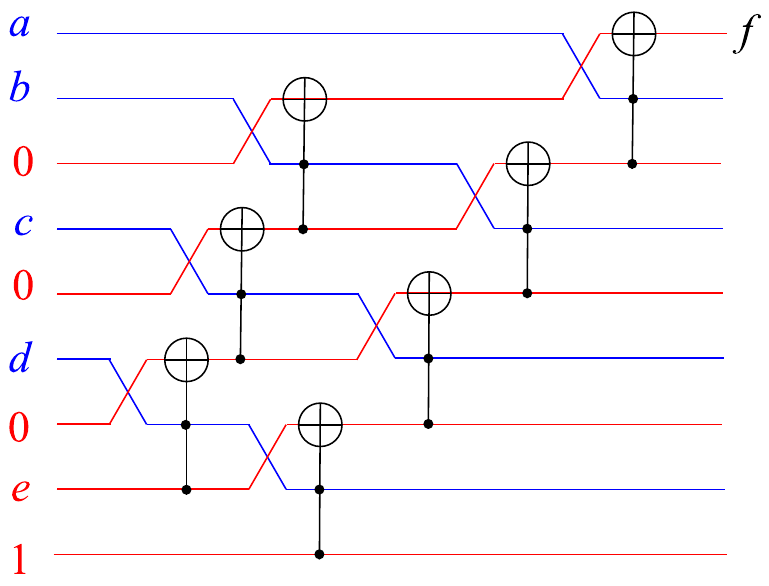}
        \caption{PDL-based circuit for $f$, where the rectangles represent SWAT gates.}
        \label{fig:example6-pdl-circuit}
    \end{subfigure}
    \begin{subfigure}[!htb]{0.25\linewidth}
        \centering
        \includegraphics[width=\linewidth]{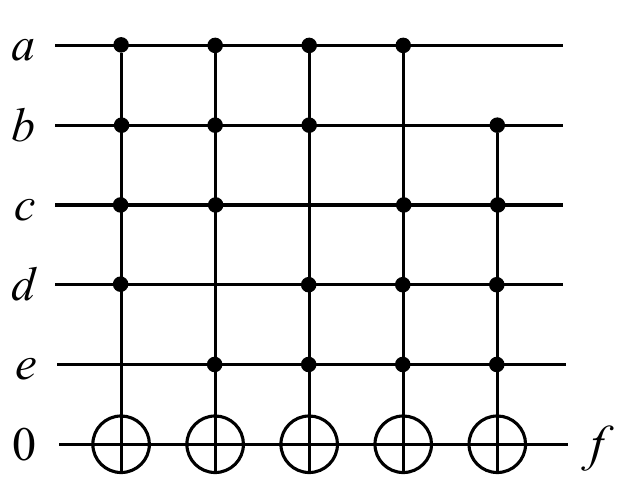}
        \caption{Circuit for $f$ realized directly from the ESOP expression where $f =  abcd \oplus abce \oplus abde \oplus acde \oplus bcde$, which can be found from the PDL.}
        \label{fig:example6-esop-circuit}
    \end{subfigure}
    \caption{Example 6 of PDL and circuits with the benchmark rd53f1 \cite{mcnc}.}
\end{figure}

\begin{figure}[!htb]
    \centering
    \begin{subfigure}[!htb]{\linewidth}
        \centering
        \includegraphics[width=0.9\linewidth]{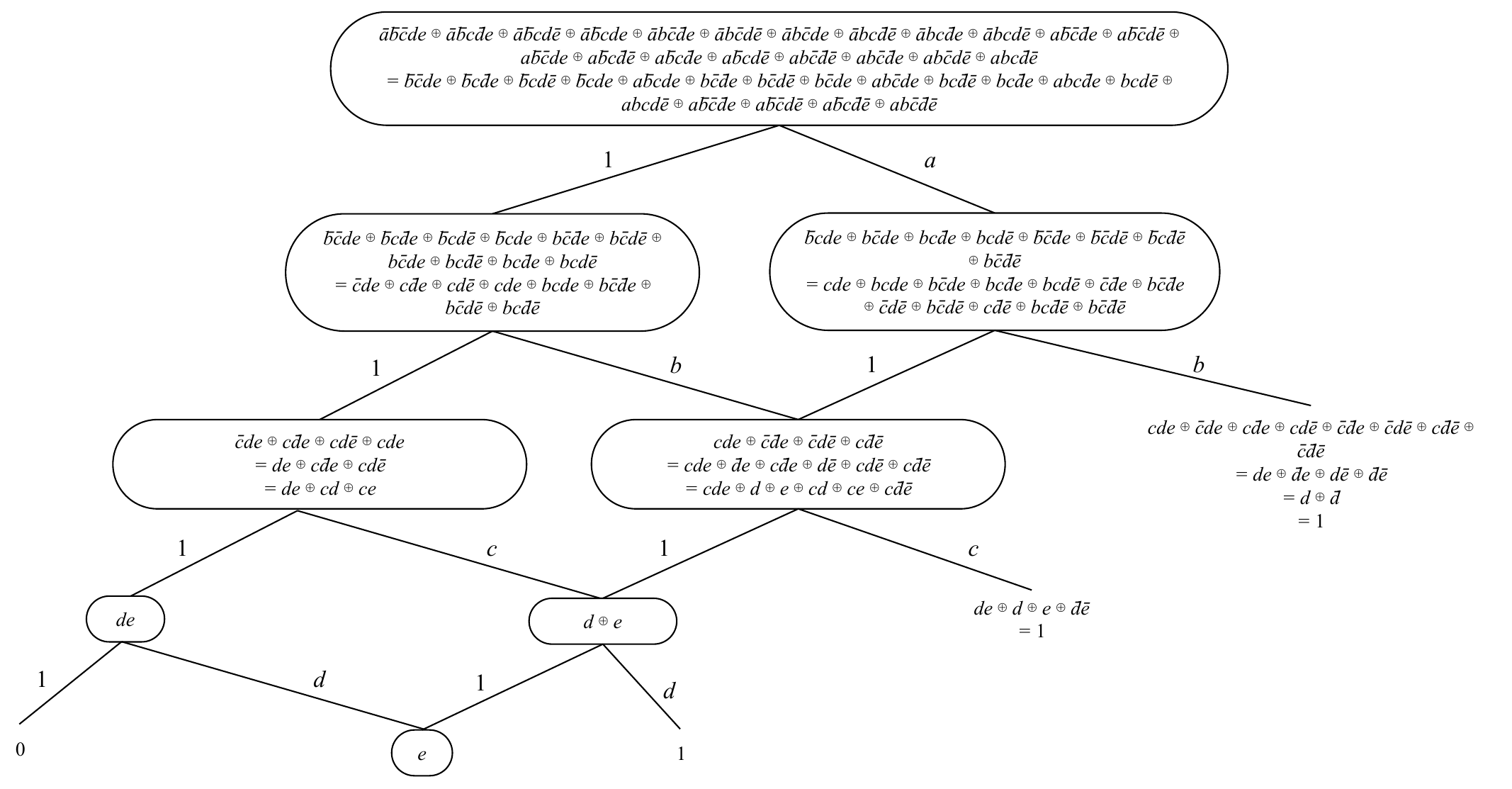}
        \caption{PDL for $f$.}
        \label{fig:example7-lattice}
    \end{subfigure}
    \begin{subfigure}[!htb]{0.35\linewidth}
        \centering
        \includegraphics[width=\linewidth]{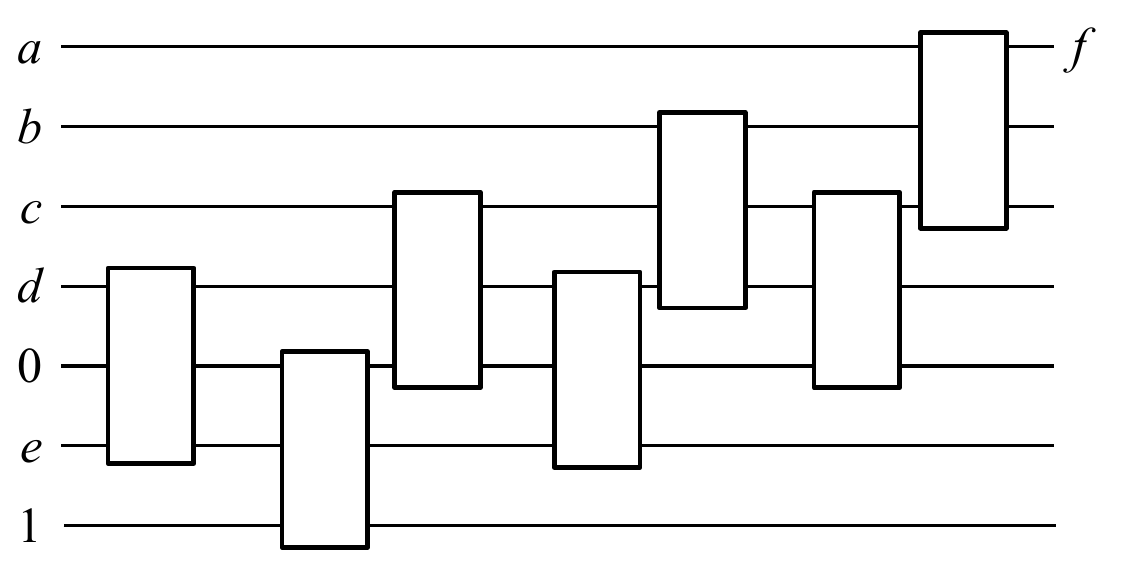}
        \caption{PDL-based circuit for $f$, where the rectangles represent SWAT gates.}
        \label{fig:example7-pdl-circuit}
    \end{subfigure}
    \begin{subfigure}[!htb]{0.5\linewidth}
        \centering
        \includegraphics[width=0.7\linewidth]{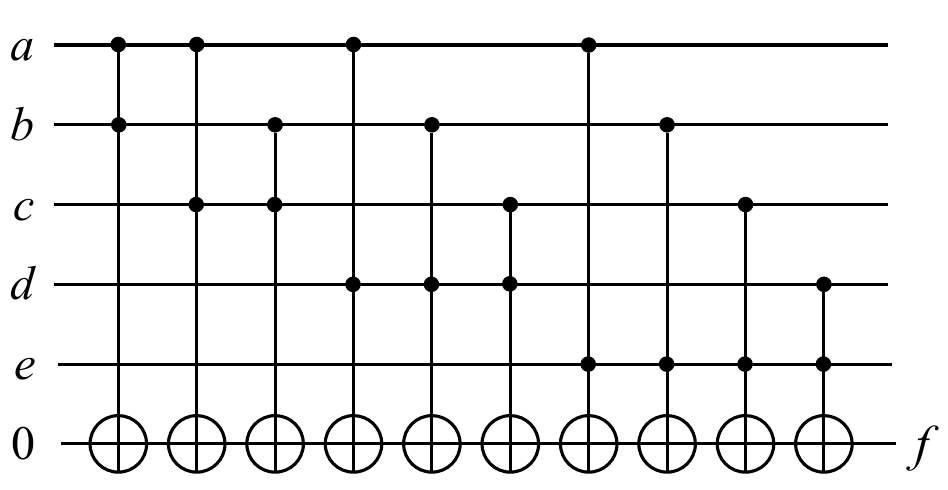}
        \caption{Circuit for $f$ realized directly from the ESOP expression, where $f = ab \oplus ac \oplus bc \oplus ad \oplus bd \oplus cd \oplus ae \oplus be \oplus ce \oplus de$, which can be found from the PDL.}
        \label{fig:example7-esop-circuit}
    \end{subfigure}
    \caption{Example 7 of PDL and circuits with the benchmark rd53f2 \cite{mcnc}.}
\end{figure}

\noindent 
\\\textbf{Comparison Example 6:} The benchmark rd53f1 \cite{mcnc} which is the function $f = \bar{a}bcde \oplus a\bar{b}cde \oplus ab\bar{c}de \oplus abc\bar{d}e \oplus abcd\bar{e} \oplus abcde$.

The PDL for $f$ is shown in Fig.~\ref{fig:example6-lattice}, and the PDL-based circuit is shown in Fig.~\ref{fig:example6-pdl-circuit}. It has a Maslov cost of 56 and TQC of 242 (when used with CALA-$n$). The ESOP-based circuit for $f$ is shown in Fig.~\ref{fig:example6-esop-circuit}, and has a Maslov cost of at least 228 (excluding additional global SWAP gates) and TQC of 2399. The PDL-based circuit synthesis method leads to a gain of at least 172 Maslov cost and 2157 TQC (with CALA-$n$).

\noindent 
\\\textbf{Comparison Example 7:} The benchmark rd53f2 \cite{mcnc} which is the function $f = \bar{a}\bar{b}\bar{c}de \oplus \bar{a}\bar{b}c\bar{d}e \oplus \bar{a}\bar{b}cd\bar{e} \oplus \bar{a}\bar{b}cde \oplus \bar{a}b\bar{c}\bar{d}e \oplus \bar{a}b\bar{c}d\bar{e} \oplus \bar{a}b\bar{c}de \oplus \bar{a}bc\bar{d}\bar{e} \oplus \bar{a}bc\bar{d}e \oplus \bar{a}bcd\bar{e} \oplus a\bar{b}\bar{c}\bar{d}e \oplus a\bar{b}\bar{c}d\bar{e} \oplus a\bar{b}\bar{c}de \oplus a\bar{b}c\bar{d}\bar{e} \oplus a\bar{b}c\bar{d}e \oplus a\bar{b}cd\bar{e} \oplus ab\bar{c}\bar{d}\bar{e} \oplus ab\bar{c}\bar{d}e \oplus ab\bar{c}d\bar{e} \oplus abc\bar{d}\bar{e}$.

The PDL and PDL-based circuit for $f$ are shown in Fig.~\ref{fig:example7-lattice} and Fig.~\ref{fig:example7-pdl-circuit}, respectively. The PDL-based circuit has a Maslov cost of 56 and TQC of 262 (when used with CALA-$n$). The ESOP-based circuit for $f$, shown in Fig.~\ref{fig:example7-esop-circuit}, has a Maslov cost of 50 (excluding additional global SWAP gates) and TQC of 711. The PDL-based circuit synthesis method leads to a gain of 449 TQC (with CALA-$n$).

The total Maslov costs and the TQC after transpilation (along with the number of each native gate) of the PDL-based circuits versus the ESOP-based circuits in the same seven comparison examples are listed in Table~\ref{tab:example-maslov-costs} and Table~\ref{tab:example-tqc-costs}, respectively. The PDL-based method performs better than the ESOP-based method, and using the CALA-$n$ decomposition for the 3-bit Toffoli gate with the PDL-based method yields the most cost-effective results.

\begin{table}[!htb]
    \centering
    \renewcommand{\arraystretch}{1.5}
    \resizebox{0.6\linewidth}{!}{
    \begin{tabular}{l||c|c}
         Example & PDL-Based & ESOP-Based \\
         \hline
         1 (4-Variable Symmetric Function) & 56 & 69  \\
         2 (5-Variable Symmetric Function) & 120 & 215 \\
         3 (5-Variable Symmetric Function) & 120 & 183 \\
         4 (11-Variable Function with Symmetric Parts) & 120 & 750  \\
         5 (6-Variable Symmetric Function) & 124 & 547 \\
         6 (rd53f1) & 56 & 145 \\
         7 (rd53f2) & 56 & 50 \\
    \end{tabular}
    }
    \caption{The Maslov costs of the different circuits in the comparison examples.}
    \label{tab:example-maslov-costs}
\end{table}

\begin{table}[!htb]
    \centering
    \renewcommand{\arraystretch}{1.5}
    \resizebox{\linewidth}{!}{
    \begin{tabular}{l||c|c|c|c|c|c||c|c|c|c|c|c||c|c|c|c|c|c}
         & \multicolumn{6}{|c||}{PDL-Based} & \multicolumn{6}{|c||}{PDL-Based with CALA-$n$ \cite{albayaty-cala}} & \multicolumn{6}{|c}{ESOP-Based} \\
         \cline{2-19}
         Ex. & $\sqrt{X}$ & $RZ$ & $CZ$ & $X$ & Depth & TQC & $\sqrt{X}$ & $RZ$ & $CZ$ & $X$ & Depth & TQC & $\sqrt{X}$ & $RZ$ & $CZ$ & $X$ & Depth & TQC \\
         \hline
         1
         & 108 & 89 & 63 & 2 & 129 & 377
         & 83 & 49 & 36 & 2 & 76 & 236
         & 332 & 162 & 161 & 3 & 444 & 1048
         \\
         2 
         & 254 & 196 & 135 & 4 & 264 & 825
         & 187 & 97 & 87 & 4 & 143 & 490
         & 1228 & 618 & 558 & 7 & 1522 & 3709 \\
         3 
         & 254 & 196 & 135 & 3 & 264 & 828
         & 187 & 97 & 87 & 3 & 143 & 489
         & 1147 & 582 & 444 & 2 & 1384 & 3331
         \\
         4 
         & 260 & 202 & 135 & 0 & 266 & 831
         & 191 & 99 & 87 & 0 & 142 & 493
         & 2960 & 1507 & 1247 & 2 & 3722 & 8876
         \\
         5 
         & 290 & 202 & 156 & 0 & 275 & 881
         & 205 & 105 & 96 & 0 & 146 & 520
         & 3509 & 1841 & 1327 & 2 & 3853 & 9776
         \\
         6
         & 109 & 90 & 63 & 1 & 129 & 378
         & 89 & 47 & 39 & 1 & 76 & 242
         & 779 & 395 & 382 & 10 & 961 & 2399
         \\
         7
         & 132 & 92 & 72 & 1 & 207 & 484
         & 80 & 48 & 36 & 1 & 107 & 262 
         & 202 & 120 & 111 & 0 & 318 & 711
    \end{tabular}
    }
    \caption{The costs of the different circuits in the comparison examples after transpilation, using the IBM Qiskit Transpiler~\cite{qiskit} with the native gates ($\sqrt{X}$, $RZ$, $CZ$, and $X$) and layout of IBM Torino quantum computer. The CALA-$n$ operator \cite{albayaty-cala} for the 3-bit Toffoli gate  (see Fig.~\ref{fig:connectivity-v2}), i.e., $n=3$, is also evaluated on the PDL-based circuits.}
    \label{tab:example-tqc-costs}
\end{table}

Our method can be effective for non-symmetric functions, but it is particularly effective for symmetric functions and those that contain a large symmetric component. This is especially the case for large, many-variable symmetric functions such as those in Examples 1 and 2, or functions with large symmetric parts, such as in Example 4. However, although this method is not effective for purely affine functions (e.g. $a \oplus b \oplus c$), it can be effective for functions containing affine functions, as shown in the comparison examples.

These comparison examples demonstrate that our PDL-based method for circuit synthesis is effective for ESOP expressions with a large number of variables and a high number of literals in many of its terms, as this method avoids $n$-bit Toffoli gates, where $n > 3$, which are shown to have significantly higher costs. In addition, these $n$-bit Toffoli gates require many more additional global SWAP gates than a 3-bit Toffoli gate. However, this method does create a large number of additional ancilla qubits.

Our introduced PDL-based circuit synthesis method utilizes only local SWAP and 3-bit Toffoli gates, leading to lower Maslov costs and fewer additional global SWAP gates.

\section{Conclusions}
This paper expands on our previous Positive Davio Lattice-based (PDL-based) circuit synthesis method \cite{lukac} and introduces new decompositions and quantum layout mappings for this method. These decomposed and mapped circuits are only constructed from SWAP and 3-bit Toffoli gates, without large $n$-bit Toffoli gates ($n \geq 3$), as these large gates, utilized by other methods, always have high quantum costs and are inefficient to map.

In this research, we introduce the ``SWAT" gate, which is equivalent to one SWAP gate followed by one 3-bit Toffoli gate. The advantage of our SWAT gate is that it is suitable for cost-effective circuit synthesis constructed from PDLs, along with qubit routing and gates mapping onto a regular quantum layout. The optimization of the SWAT gate is a potential future research topic. To distinguish from other SWAP gates outside a SWAT gate, we defined the outside SWAP gates as ``global" and the inside SWAP gates as ``local".

By studying contemporary quantum layouts, the square grid layout has four neighboring qubits for each qubit \cite{arute, helmer, gong}, and the heavy-hex layout has two or three neighboring qubits for qubit \cite{chamberland}. Due to these physical connectivity limitations for various decompositions, we introduce a new quantum ``triangular" layout. This triangular layout has six neighboring qubits for each qubit, and is suitable for our PDL-based circuit synthesis method since this layout does not require any additional SWAP gates. Although this triangular layout has not yet been fabricated in a real quantum computer, as quantum technology continues to advance, this layout could become a powerful architecture in the future.

In addition, we introduce mappings of the connectivity among the physical qubits in a PDL-based circuit for the square grid and heavy-hex layouts. These mappings do not require additional global SWAP gates other than those already a part of the SWAT gates, leading to cost-effective circuits.

To systematically test these PDL-based circuits, we investigated seven comparison examples, which demonstrate that these circuits generally have a lower Maslov cost and a lower Transpilation Quantum Cost (TQC) than those directly built from an Exclusive-or Sum of Products (ESOP) expression. This is especially the case for symmetric functions and ESOP expressions that would otherwise require a high number of large, cost-expensive $n$-bit Toffoli gates ($n>3$) and global SWAP gates.

In conclusion, there are many potential future directions. The triangular layout could be expanded into 3-dimensional architectures, a promising path for quantum layouts. Ways of reducing noise in this layout can also be studied. As for our PDL-based circuit synthesis method, similar methods for expanding to multi-output functions and with lattices based on other expansions or a combination of multiple expansions have yet to be intensively studied and evaluated.

\nonumsection{References}

\end{document}